\documentclass[fleqn,usenatbib]{mnras}

\usepackage{newtxtext,newtxmath}

\usepackage[T1]{fontenc}
\usepackage{multirow}
\usepackage{multicol}
\usepackage{booktabs,rotating}
\usepackage{lscape}
\usepackage{xcolor}

\DeclareRobustCommand{\VAN}[3]{#2}
\let\VANthebibliography\thebibliography
\def\thebibliography{\DeclareRobustCommand{\VAN}[3]{##3}\VANthebibliography}

\usepackage{graphicx}	
\usepackage{amsmath}	

\usepackage{amssymb}

\usepackage{ulem}

\title[Eclipse timing variations in RR Cae]{Eclipse timing variations in the WD+dM eclipsing binary RR Cae}

\author[R. Rattanamala et al.]{R. Rattanamala,$^{1,2}$
S. Awiphan,$^{3}$\thanks{E-mail: supachai@narit.or.th}
S. Komonjinda,$^{4}$
A. Phriksee,$^{3}$
P. Sappankum,$^{1,3}$ \newauthor
N. A-thano,$^{5}$ 
S. Chitchak$,^{6}$
P. Rittipruk,$^{3}$
U. Sawangwit,$^{3}$
S. Poshyachinda, $^{3}$  \newauthor
D. E. Reichart, $^{7}$ 
and
J. B. Haislip $^{7}$
\\
$^{1}$PhD Program in Astronomy, Department of Physics and Materials Science, Faculty of Science, Chiang Mai University, Chiang Mai, 50200, Thailand\\
$^{2}$Physics and General Science Program, Faculty of Science and Technology, Nakhon Ratchasima Rajabhat University, Nakhon Ratchasima 30000, Thailand\\
$^{3}$National Astronomical Research Institute of Thailand (Public Organization), 260 Moo 4, Donkaew, Mae Rim, Chiang Mai, 50180, Thailand\\
$^{4}$Department of Physics and Materials Science, Faculty of Science, Chiang Mai University, Chiang Mai, 50200, Thailand\\
$^{5}$Institute of Astronomy, National Tsing Hua University, Hsinchu 30013, Taiwan\\
$^{6}$Department of Physics, Faculty of Science, Udon Thani Rajabhat University, Udon Thani 41000, Thailand\\
$^{7}$Department of Physics and Astronomy, University of North Carolina at Chapel Hill, Chapel Hill NC 27599, USA \\
}

\date{Accepted XXX. Received YYY; in original form ZZZ}

\pubyear{2021}

\usepackage{float}
\usepackage{breqn}

\begin{document}
\label{firstpage}
\pagerange{\pageref{firstpage}--\pageref{lastpage}}
\maketitle

\begin{abstract}

We present the binary model and the eclipse timing variations of the eclipsing binary RR Cae, which consists of a white dwarf eclipsed by an M-type dwarf companion. The multi-wavelength optical photometry from the Transiting Exoplanet Survey Satellite (TESS), the 0.6-m PROMPT-8 telescope, and the 0.7-m Thai Robotic Telescope at Spring Brook Observatory, combined with archive H-alpha radial velocities from the Very Large Telescope (VLT) are analysed. From the data, the physical parameters of the system are obtained along with 430 new times of minima. The TESS light curves in 2018 and 2020 show out-of-eclipse variations, which might be caused by a large spot on the secondary component. The light travel time effect models due to the gravitational interaction of one or two circumbinary objects are adopted to fit the cyclic variations in the RR Cae’s O-C curve. The fitting solution of the O-C curve with one circumbinary object model shows a periodic variation with a period of $16.6\pm0.2$~yr and an amplitude of $14\pm1$ s, which can be caused by a planet with a minimum mass of $3.4\pm0.2$~M$_{\textup{Jup}}$. When we consider the model with two circumbinary objects, the O-C curve shows cyclic variations with periods of $15.0\pm0.5$ yr and $39\pm5$ yr and amplitudes of $12\pm1$ s and $20\pm5$ s, respectively, corresponding to minimum masses of $3.0\pm0.3$ M$_{\textup{Jup}}$ and $2.7\pm0.7$ M$_{\textup{Jup}}$.

\end{abstract}




\section{Introduction}
Eclipsing binaries are important for analysing the physical parameters and evolution of stars. Post-Common Envelope Binaries (PCEBs) are pre-cataclysmic variables generally consisting of a white dwarf and a main-sequence star. As the main sequence component transfers mass to the white dwarf companion, it is the evolutionary phase of binaries before they evolve into cataclysmic variables (CVs) \citep{Warner1995}. In this evolutionary phase, the system might lose orbital angular momentum via magnetic braking from the main-sequence star and tidal coupling \citep{Ribeiro2013}.

The PCEB primary component is either a sub-dwarf O/B (sdOB) or a white dwarf (WD), while the secondary component is a main-sequence (MS). The significant differences between effective temperatures and radii of the hot primary and the cold secondary produce deep primary eclipses, which can be used to create the O-C diagrams accurately. The O-C diagram can be used to study eclipsing binaries' evolution and search for circumbinary brown dwarfs or planets (e.g. HW Vir \citep{Beuermann2012}, NY Vir \citep{Lee2014}, V2051 Oph \citep{Qian2015}, DE CVn \citep{Han2018}).

Nowadays, more than 5000 exoplanets have been discovered, including over a hundred planets in binary systems, circumbinary (orbiting both stars), or circumstellar (orbiting only one star) configurations\footnote{The Extrasolar Planets Encyclopedia: \texttt{http://exoplanet.eu/}}. Most circumbinary planets were discovered via transit (e.g. Kepler-16 b \citep{Doyle2011}, Kepler-1647~b \citep{Kostov2016}) or eclipsing binary timing techniques (e.g. HW~Vir~b \citep{Beuermann2012}, DP~Leo~b \citep{Qian2010}). These circumbinary planets can provide a better understanding of the planetary formation and evolution, especially planets around double-lined eclipsing binary for which precise radii and masses of the stellar components can be obtained \citep{Qian2012,Borkovits2013}.

\begin{table*}
    \begin{center}
	\caption{Summarized of the RR Cae observation data.}
	\label{tab:LC_log}
	\begin{tabular}{cccccc}
		\hline \hline
		\multirow{2}{*}{Telescope} & \multirow{2}{*}{Observation date} & Sampling time/ & \multirow{2}{*}{Filters} & \multirow{2}{*}{Numbers of data} & \multirow{2}{*}{Numbers of eclipses} \\
		& &Exposure time (s)\\
		\hline
		TESS         & 20/9/2018 -- 16/12/2020 &  120$^*$ & TESS bandpass & 88 881   & 406 \\
		PROMPT-8     & 22/9/2018 -- 06/10/2018 &  120$^\dagger$ & R    & 199      & 4 \\
		TRT-SBO      & 4/10/2019 -- 12/12/2020 &  30$^\dagger$  & R    & 5373    & 14 \\
		TRT-SBO      & 1/11/2020 -- 22/12/2020 &  30$^\dagger$  & I    & 2057    & 6 \\
		\hline
	\end{tabular}
	\end{center}
	\begin{flushleft}
	\textbf{Notes.} $^*$ represents sampling time. $^\dagger$ represents exposure time. \\
	\end{flushleft}
\end{table*}

\begin{figure*}
	\includegraphics[width=2\columnwidth]{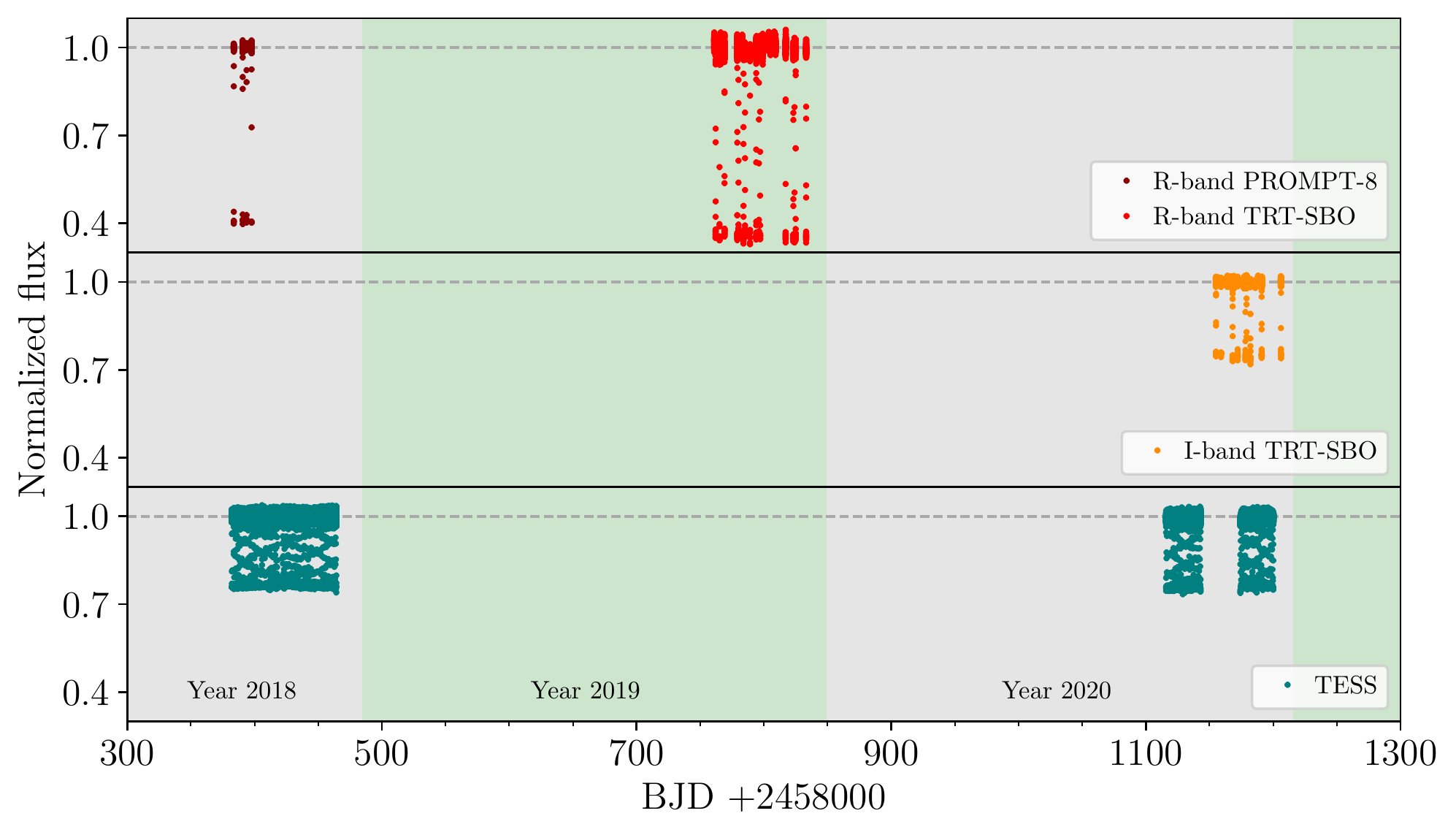}
    \caption{The RR Cae light curves. The Upper panel: light curves in the R-band from the PROMPT-8 telescope (Dark red) and the TRT-SBO (Red). The middle panel: light curves in the I-band from the TRT-SBO (Orange). The lower panel: light curves from the TESS (Teal). Dashed lines show a normalised baseline.}
    \label{fig:LC}
\end{figure*}

The eclipsing binary system RR Cae ($\alpha_{2000}$ = $04^{\textup{h}}21^{\textup{m}}5^{\textup{s}}.56$, $\delta_{2000}$ = -$48^{\circ}39'7''.06$, G = 13.67, distance = 21.19 pc\footnote{Obtained from Gaia EDR3 catalogue from the Gaia archive website: \texttt{https://archives.esac.esa.int/gaia}}) is a detached binary discovered by \citet{Krzeminski1984}, which consists of a cool white dwarf and a late-type main-sequence star (DA7.8+M4) with an orbital period of 7.289 h \citep{Bruch1999}. \citet{BruchandDiaz1998} provided the first photometric and spectrum report and found that the primary and the secondary masses were 0.356~M$_{\odot}$ and 0.089~M$_{\odot}$, respectively. \citet{Bruch1999} measured the radial velocities of the secondary and obtained the primary mass and the secondary mass of 0.467~M$_{\odot}$ and 0.095~M$_{\odot}$, respectively. He also found a large radius of 0.189~R$_{\odot}$ for the secondary star, which was interpreted as a possible consequence of the previous common envelope phase.

\citet{Maxted2007} investigated the period change of RR Cae and reported that its orbital period did not change by more than $\left | \dot{P} \right |/P \simeq 5\times10^{-12}$ over a time scale of a decade. In 2010, \citet{Parsons2010} found that the RR Cae orbital period variation has a sinusoidal variation in the O-C diagram, which was confirmed later by \citet{Qian2012}. \citet{Qian2012} showed that the system's orbital period increased at a rate of $4.18\pm0.20~\times10^{-12}$~s/s. The O-C diagram had a periodical variation due to the third body component with a period of 11.9 yr and an amplitude of 14.3 s. The third body component was suggested to be a circumbinary planet with a mass of 4.2~$\pm$~0.4 M$_{\textup{Jup}}$ and an orbital semi-major axis of 5.3~$\pm$~0.6 AU. From the spectroscopic study of \citet{Ribeiro2013}, it was possible to determine the effective temperature and the surface gravity of the white dwarf primary ($T_{\textup{eff}}$~=~7260~$\pm$~250~K, $\log~g$~=~7.8~$\pm$~0.1~dex). Moreover, the presence of metal lines in the WD spectrum was interpreted as a signature of accretion of the M-dwarf wind onto the surface of the white dwarf, with a mass accretion rate of $(7 \pm 2) \times 10^{-16}$ M$_\odot$/yr.

In 2018, NASA’s Transiting Exoplanet Survey Satellite (TESS) was launched in order to monitor nearby bright stars \citep{TESS}. The TESS provides long-term photometric observations of RR Cae during 2018-2020. These data can be used to obtain an accurate sinusoidal variation model of the RR Cae's O-C diagram, which leads to the precise measurements of the mass and the orbital parameters of a circumbinary object. Moreover, combining the TESS data with photometric data obtained from the Thai Robotic Telescope Network (TRTN) and the published spectroscopic data from the Ultraviolet and Visual Echelle Spectrograph (UVES) at the Very Large Telescope (VLT) (Program ID: 076.D-0142, PI: Maxted, Pierre), the physical parameters of the binary can be revised.

In this paper, the physical parameters and orbital period variation analyses from additional RR Cae observations are presented. In Section~\ref{sec:Observation}, the details of observation and data analyses are reported. The physical models of the RR Cae components are shown in Section~\ref{sec:Parameters}. In Section~\ref{sec:OC}, the time of minima of RR Cae are used to analyse timing variations to obtain the new ephemeris and confirm the third body in the system. Finally, Section~\ref{sec:Conclusion} concludes this study.

\section{Observation and data analysis}
\label{sec:Observation}

\subsection{Photometric observation}

In this work, the photometric data of RR Cae obtained from three telescopes: the TESS, the PROMPT-8 telescope, and the Thai Robotic Telescope at Spring Brook Observatory (TRT-SBO) are presented. The observations were conducted between 2018 and 2020. A total of 430 primary eclipses were obtained from 96 510 data points. The photometric data are shown in Tables~\ref{tab:LC_log}, \ref{tab:data} and Figure~\ref{fig:LC}.

\subsubsection{Transiting Exoplanet Survey Satellite (TESS)}
The TESS has monitored RR Cae in Sectors 3, 4, 5, 30, and 32 between 2018 and 2020. The calibrated data were obtained from Mikulski Archive for Space Telescopes\footnote{Mikulski Archive for Space Telescopes: \texttt{https://mast.stsci.edu/}}. The sampling time of an image was 120 s in the TESS bandpass (600-1000 nm). The out-of-eclipse data above three standard deviations from the mean are cut. After the cut, the 88 881 data are used for the analyses.

\subsubsection{PROMPT-8 telescope}
\label{sec:PROMPT8}
We observed RR Cae with the CCD camera array size $2048~\times~2048$ pixels with a scale of 0.624 arcsec/pixel on the PROMPT-8 telescope, a 0.6-m robotic telescope at Cerro Tololo Inter-American Observatory (CTIO), Chile, in the R-filter between 2018 September and 2018 October. The exposure time was set to 120 s. Four primary eclipses from 199 images were obtained in total. We used the \texttt{IRAF} package, Astrometry.net \citep{Lang2010}, and \texttt{Sextractor} \citep{Bertin1996} for reduction, astrometric calibration, and photometry of the images. The light curves were obtained using aperture photometry with an adaptive scaled aperture based on the seeing of each image.

\subsubsection{Thai Robotic Telescope at Spring Brook Observatory (TRT-SBO)}
The TRT-SBO is a 0.7-m telescope located at Spring Brook Observatory, Australia. The telescope is part of the Thai Robotic Telescope Network operated by the National Astronomical Research Institute of Thailand (NARIT). RR Cae was observed with the $4096~\times~4096$~pixels ProLine PL16803 Monochrome CCD camera on the 0.7-m telescope in the R-band between 2019 October and 2019 November, and the I-band between 2020 October and 2020 November with an exposure time of 30 s for both filters. In total, 5373 images covered 14 primary eclipses in the R-band, and 2057 images covered six primary eclipses in the I-band. In order to obtain the light curves, the same procedure as the PROMPT-8 telescope was performed.

In Figure~\ref{fig:LC}, the eclipse depths in the R-band are deeper than those in the I-band, as expected. As a result, the eclipse timing in the R-band is more precise than in the I-band. The multi-waveband ground-based observations were initially designed for the stellar temperature modelling of both components. However, the TRT-SBO along with the PROMPT-8 data (Section~\ref{sec:PROMPT8}) were not used to model the physical parameters as designed due to the data quality and complexity of the starspot modelling (See Section~\ref{sec:Parameters}).

\begin{table}
	\caption{The photometric data of RR Cae from the PROMPT-8 telescope and the TRT-SBO.}
	\label{tab:data}
	\begin{tabular}{ccccc}
		\hline \hline
		 BJD & Normalised flux & Error & Filters & Observatory\\
		\hline
		2 458 383.664 10 &  0.996 &  0.003 &   R & PROMPT-8 \\
		2 458 383.665 66 &  1.009 & 0.003  &   R & PROMPT-8 \\
		2 458 383.667 25 &  1.011 & 0.003  &   R & PROMPT-8 \\
		2 458 383.668 82 &  1.013 & 0.003  &   R & PROMPT-8 \\
		$\dots$ & $\dots$ &  $\dots$ & $\dots$ & $\dots$ \\
		$\dots$ & $\dots$ &  $\dots$ & $\dots$ & $\dots$ \\
		$\dots$ & $\dots$ &  $\dots$ & $\dots$ & $\dots$ \\
		2 458 761.020 08 &  1.006 & 0.013  &   R & TRT-SBO \\
		2 458 761.020 57 &  0.989 & 0.013  &   R & TRT-SBO \\
		2 458 761.026 21 &  1.003 & 0.014  &   R & TRT-SBO \\
		2 458 761.026 68 &  1.004 &  0.013 &   R & TRT-SBO \\
		$\dots$ & $\dots$ &  $\dots$ & $\dots$ & $\dots$ \\
		$\dots$ & $\dots$ &  $\dots$ & $\dots$ & $\dots$ \\
		$\dots$ & $\dots$ &  $\dots$ & $\dots$ & $\dots$ \\
		2 459 155.033 62 &  0.998 & 0.006  &   I & TRT-SBO \\
		2 459 155.034 15 &  1.013 & 0.006  &   I & TRT-SBO \\
		2 459 155.034 76 &  1.005 & 0.006  &   I & TRT-SBO \\
		2 459 155.035 25 &  1.016 & 0.006  &   I & TRT-SBO \\
		$\dots$ & $\dots$ &  $\dots$ & $\dots$ & $\dots$ \\
		$\dots$ & $\dots$ &  $\dots$ & $\dots$ & $\dots$ \\
		$\dots$ & $\dots$ &  $\dots$ & $\dots$ & $\dots$ \\
		\hline
	\end{tabular}
	\textbf{Note.} The full table is available in electronic format in the full form.
\end{table}

\subsection{Spectroscopic data}
\label{sec:Spectrum}
Between 2005 October 15$^{\textup{th}}$ and 2005 November 16$^{\textup{th}}$, RR Cae was observed with the UVES located at the Nasmyth B focus of UT2 of the VLT (Program ID: 076.D-0142, PI: Maxted, Pierre). The 128 published spectra are obtained from the European Southern Observatory (ESO) Science Archive Facility\footnote{\texttt{http://archive.eso.org/scienceportal/home}}. Eight observed spectra with exposure times of 340~s were combined for each published spectrum. The data provide the spectral resolution of 21 000 covered wavelengths between 328.1 nm and 668.8 nm with a signal-to-noise ratio (S/N) between 15 and 64. The spectra were analysed by \citep{Ribeiro2013}. However, they have not published their radial velocity measurements. Therefore, in order to obtain the radial velocity values, we reanalysed their spectra in this work.

\begin{figure*}
	\includegraphics[width=2\columnwidth]{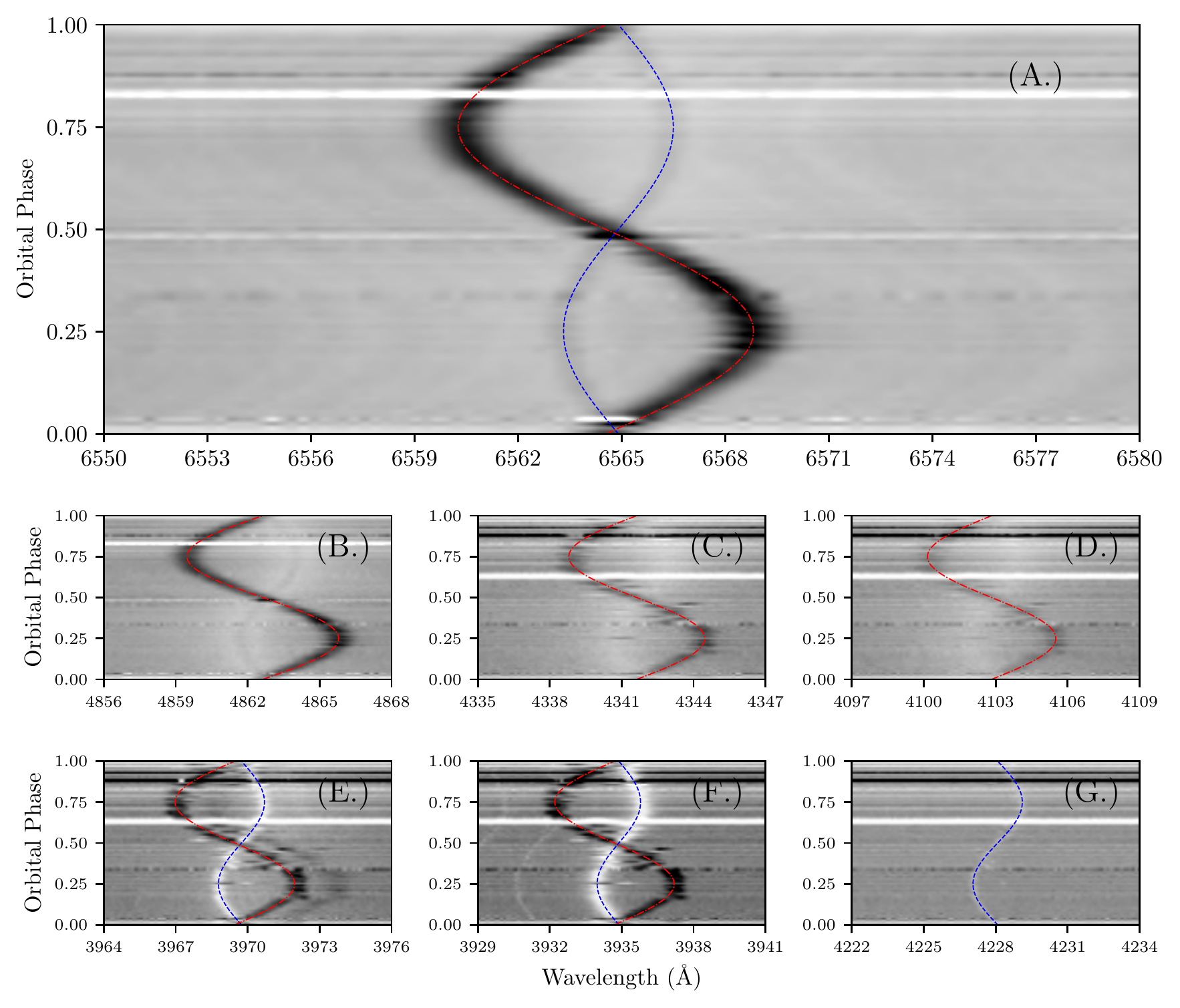}
    \caption{The spectrum lines (A.) H$_\alpha$, (B.) H$_\beta$, (C.) H$_\gamma$, (D.) H$_\delta$, (E.) Ca II (H line), (F.) Ca II (K line) and (G.) Ca I of RR Cae. The blue dashed lines show the fitting of the primary component spectrum, and the red dashed-dot lines show the fitting of the secondary component spectrum (See Section~\ref{sec:Spectrum}). The spectra are phase-binned with 0.01 phase resolution. For visualisation, the images are performed bicubic interpolation.} The orbital phases used in these figures are calculated from the ephemeris of \citet{Maxted2007}.
    \label{fig:Spectrum}
\end{figure*}

From the published spectra, both emission and absorption lines of Ca II (H line, 3968~{\AA}) and Ca II (K line, 3934~{\AA}) are detected with the emission lines of H$_\alpha$ (6563~{\AA}), H$_\beta$ (4861~{\AA}), H$_\gamma$ (4340~{\AA}), H$_\delta$ (4102~{\AA}) and absorption line of Ca I (4227~{\AA}). The published spectra were performed barycentric correction. The phase curves of the spectra are shown in Figure~\ref{fig:Spectrum}. The phases of the system were calculated from the linear ephemeris equation of \citet{Maxted2007},

\begin{equation}
    \mbox{Min.} \mbox{I} = \mbox{BJD}\ 2\ 451\ 523.048\ 567 + 0.303\ 703\ 6366\  \mbox{d} \times E \ ,
	\label{eq:ephemeris}
\end{equation}
where $E$ is an observed epoch. In order to find the radial velocities of the emission and absorption lines, the emission and absorption lines are individually fitted by the Gaussian function using the \texttt{Scipy} Optimize Curve Fit package. The radial velocities of the spectral lines with their uncertainty calculated from their covariance matrices are shown in Table~\ref{tab:RV_results}.

In order to find the semi-amplitude ($K$) and the systemic velocity ($\gamma$), the system is assumed to be in a circular orbit. The fitted semi-amplitudes of each emission or absorption line were performed using the Optimize Curve Fit package with weighted error, which determined the standard deviations of errors. Their fitting results are shown in Figure~\ref{fig:Spectrum} and Table~\ref{tab:RV_KGam}. We obtained the average semi-amplitudes of 73.4~$\pm$~0.5~km~s$^{-1}$ for the primary component and 196.0~$\pm$~1.0~km~s$^{-1}$ for the secondary component. The obtained semi-amplitude of the primary component is smaller than the semi-amplitude of 79.3~$\pm$~3.0~km~s$^{-1}$ obtained by \citet{Maxted2007}, which is combined from H$_\beta$, H$_\gamma$, and H$_\delta$ lines. However, the obtained value is compatible with the semi-amplitude value of \citet{Ribeiro2013}, 74.1~$\pm$~2.3~km~s$^{-1}$, which used the average of the semi-amplitude values of H-alpha and H-beta lines. Due to weak H$_\beta$ emission of the primary component, their uncertainty in the semi-amplitude is larger than ours. For the secondary component, \citet{Maxted2007} computed the cross-correlation in the region between 8440~\AA~and 8930~\AA~and obtained the semi-amplitude of the secondary component, which is 190.2~$\pm$~3.5~km~s$^{-1}$. In comparison, \citet{Ribeiro2013} obtained the semi-amplitude of the secondary component of 194.4~$\pm$~0.7~km~s$^{-1}$. Both values are compatible with our obtained average semi-amplitude of 196.0~$\pm$~1.0~km~s$^{-1}$.

In this work, there are only three spectral lines: H$_\alpha$, Ca II (K line), and Ca II (H line), which can be used to calculate radial velocity solutions for both primary and secondary components shown in Figure~\ref{fig:Spectrum}. From these three spectral lines, the average mass ratio of the RR Cae system, $q~=~\overline{K_{\textup{primary}}} / \overline{K_{\textup{secondary}}}$, is 0.374~$\pm$~0.001. The mass ratio is smaller than the ratio of $\sim$0.42 computed by \citet{Maxted2007}, but the calculated ratio is compatible with the ratio of \citet{Ribeiro2013}, 0.376~$\pm$~0.005, which used the same dataset.

\begin{figure}
	\includegraphics[width=\columnwidth]{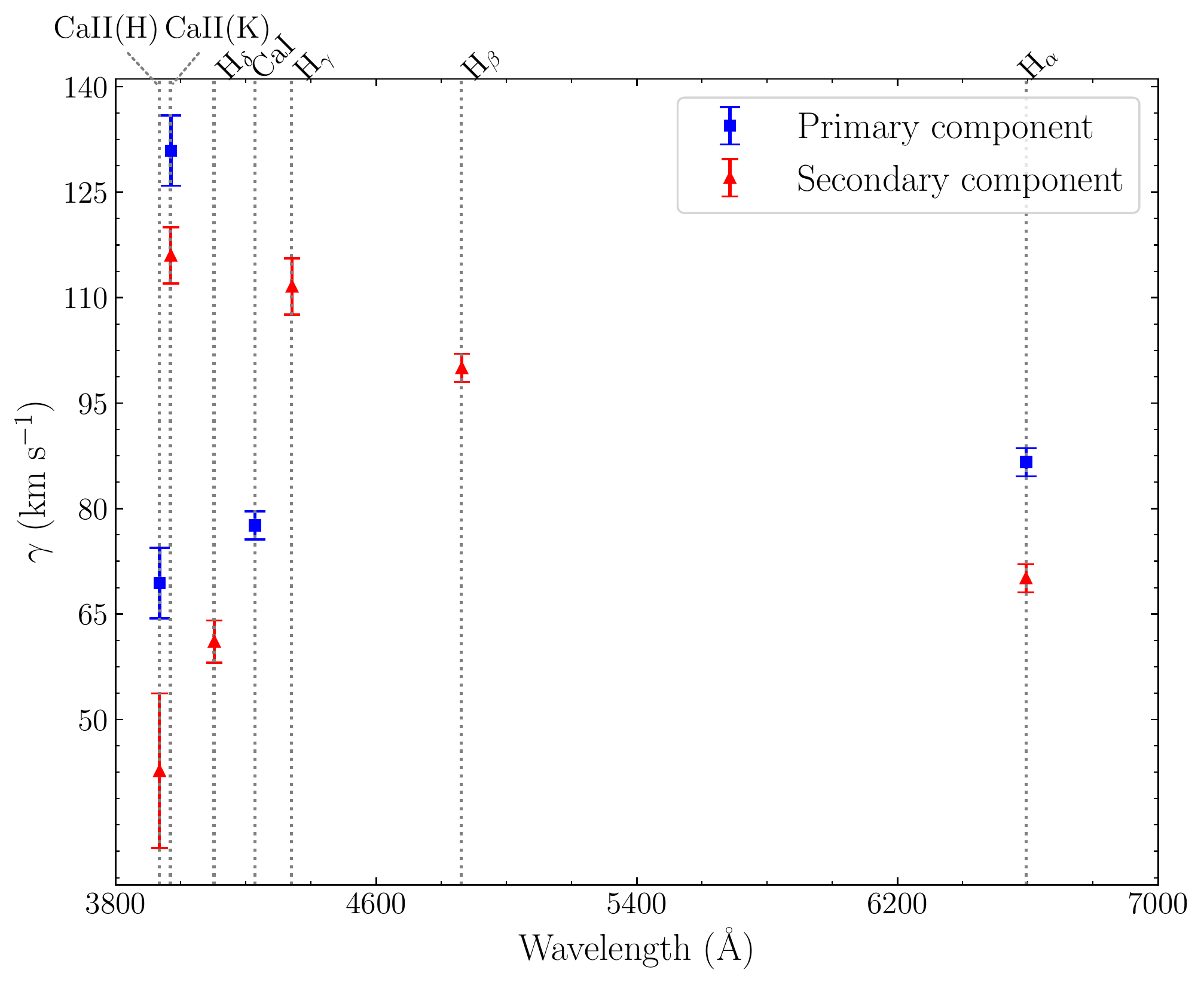}
    \caption{The non-relationships between wavelengths and systemic velocities of each spectrum feature of the primary (Blue square) and the secondary (Red triangle) components. The error bars are shown at ten times their actual size.}
    \label{fig:WaveGamma}
\end{figure}

The radial velocity solutions of the systemic velocities show variations. However, there is no relation between the variations and the spectral wavelength, as shown in Figure~\ref{fig:WaveGamma}. As the difference in the systemic velocity of the primary and the secondary components ($\gamma_{\textup{primary}}~\neq~\gamma_{\textup{secondary}}$), \citet{Maxted2004} suggested that three factors might cause this difference:

\begin{itemize}
    \item The gravitational redshift effect from the primary component. \citet{Ribeiro2013} suggested that the effect is the most reasonable explanation for the difference in systemic velocity. They found that the gravitational redshift of the RR Cae system is about $17.2~\pm~0.5$~km~s$^{-1}$. While in this work, the average gravitational redshift of three aforementioned lines: H$_\alpha$, Ca II (K line) and Ca II (H line), is around $16.2~\pm~0.4$~km~s$^{-1}$.
    \item The Stark effect is asymmetrical for the higher Balmer lines resulting in pressure shifts \citep{grabowski1987}.
    \item The tilt in the continuum will result in a systematic error in the value of the systemic velocity of the white dwarf measured from such broad lines \citep{Maxted2004}.
\end{itemize}

The variations might also be caused by the secondary star heated by the white dwarf and vice versa. The heated surface of the heated star, its inward-facing hemisphere, is hotter than the outward hemisphere. Therefore, the centre-of-light of the system is not the same as the system's centre-of-mass.

In order to eliminate systemic velocity variations, only the H$_\alpha$ spectral lines are used in this work as the highest signal-to-noise spectral lines. In total, 91 and 128 high signal-to-noise H$_\alpha$ radial velocities of the primary and secondary components are used to model the physical parameters of the RR Cae system in Section~\ref{sec:Parameters}.

\begin{figure}
	\includegraphics[width=1\columnwidth]{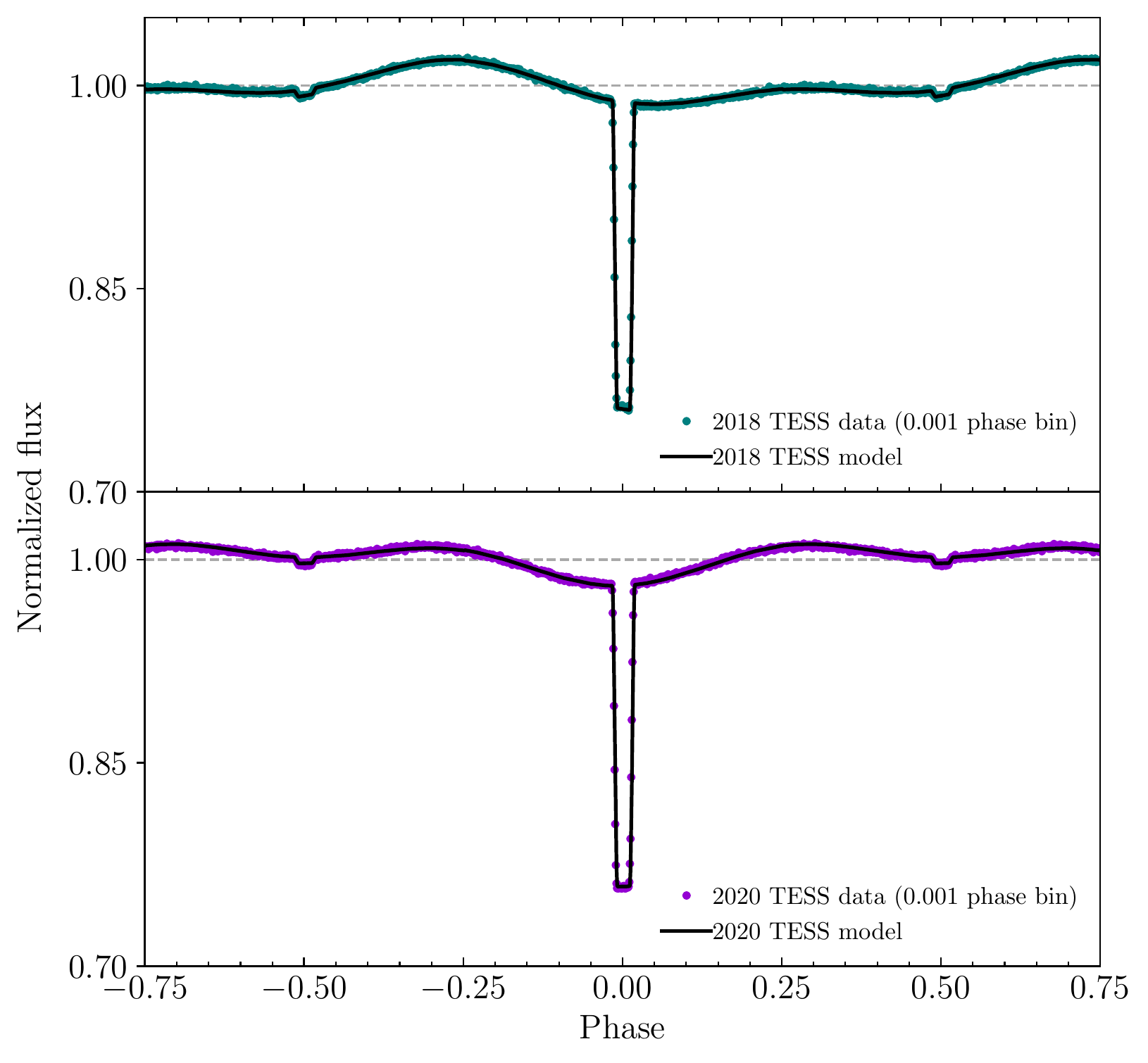}
    \caption{The RR Cae phase-folded light curves were obtained from the TESS in 2018 (Teal) and 2020 (Purple) with the best-fitting models from the \texttt{PHOEBE} code (Black solid lines). The dashed lines show normalised baselines.}
    \label{fig:LC_TESS}
\end{figure}

\begin{figure}
	\includegraphics[width=1\columnwidth]{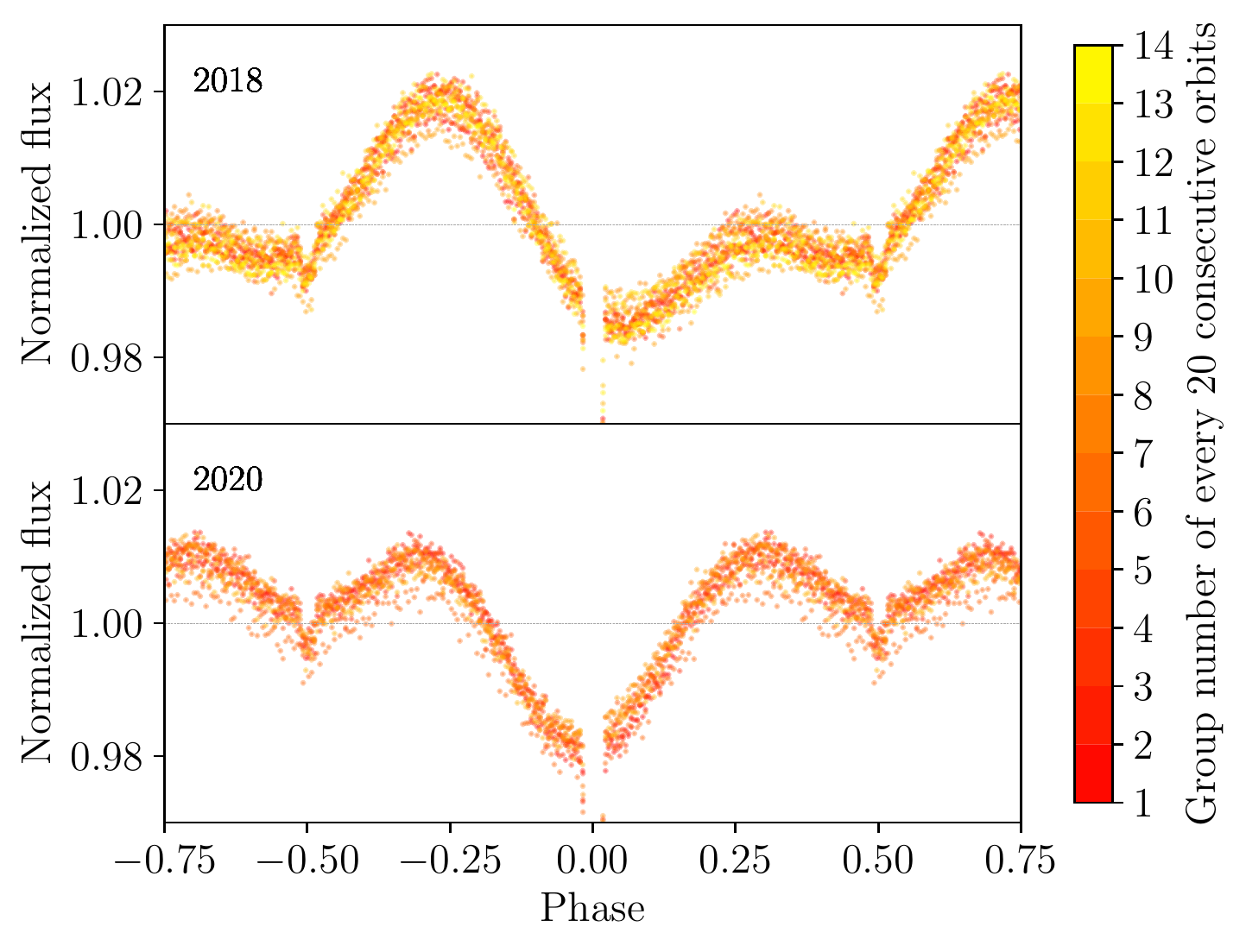}
    \caption{The average TESS normalised flux of every 20 consecutive orbits ($\sim$6~d) in 2018 (Top) and 2020 (Bottom). The light curves are phase-binned with 0.005 phase resolution. The colour scale indicates the group number of every 20 consecutive orbits in chronological order for each year.}
	\label{fig:LC_TESS_Group}
\end{figure}

\begin{table}
	\centering
	\caption{The semi-amplitude ($K$) and the systemic velocity ($\gamma$) from each spectra line measurements.}
	\label{tab:RV_KGam}
	\begin{tabular}{lcrr}
		\hline \hline
		Features & Wavelength ({\AA}) & $K$ (km s$^{-1}$) & $\gamma$ (km s$^{-1}$) \\
		\hline \hline
		\multicolumn{4}{|c|}{\emph{Primary component}} \\
		\hline
		H$_\alpha$       & 6594.9 & 73.2 $\pm$ 0.3  & 86.6 $\pm$ 0.2 \\
		Ca II (K line)   & 3934.9 & 73.8 $\pm$ 0.8  & 130.9 $\pm$ 0.5 \\
		Ca II (H line)   & 3969.7 & 73.6 $\pm$ 0.7  & 69.4 $\pm$ 0.5 \\
		Ca I  (4227{\AA})& 4228.1 & 72.8 $\pm$ 0.3  & 77.6 $\pm$ 0.2 \\
		\hline \hline
		\multicolumn{4}{|c|}{\emph{Secondary component}}\\
		\hline
		H$_\alpha$    & 6594.5 & 195.4 $\pm$ 0.3   & 70.1 $\pm$ 0.2 \\
		H$_\beta$     & 4862.6 & 196.5 $\pm$ 0.4   & 100.0 $\pm$ 0.2 \\
		H$_\gamma$    & 4341.6 & 195.6 $\pm$ 0.5   & 61.1 $\pm$ 0.3 \\
		H$_\delta$    & 4102.8 & 196.8 $\pm$ 0.6   & 111.6 $\pm$ 0.4 \\
		Ca II (K line)& 3934.7 & 197.4 $\pm$ 0.6   & 116.0 $\pm$ 0.4 \\
		Ca II (H line)& 3969.5 & 194.8 $\pm$ 2.8   & 42.7 $\pm$ 1.1 \\
		\hline
    \end{tabular}
\end{table}

\begin{table}
	\centering
	\caption{The prior range for the MCMC fitting of the RR Cae TESS light curves.}
	\label{tab:prior}
	\begin{tabular}{lc}
		\hline \hline
	    Parameters  & Prior range\\
		\hline
	    $t_{0}$ (BJD-2 450 000) &  1523.04 -- 1523.06 \\
	    $P_{0}$ &  0.303 700 -- 0.303 708   \\
	    $i$ &  75 -- 90   \\
	    $a$ ${(R_\odot)}$ &   1.5 -- 1.7    \\
	    $T_{2}$ (K) &  2000 -- 6000   \\
	    $q$ & 0.30 -- 0.45    \\
	    $R_{1}$ ${(R_\odot)}$ &  0.0001 -- 0.1   \\
	    $R_{2}$ ${(R_\odot)}$ &  0.15 -- 0.40   \\
	    \hline
		\multicolumn{2}{|c|}{\emph{Light curve parameters}}\\
		\hline
	    $F_{\textup{background,TESS2018}}$ & 0.0 -- 0.5      \\
	    $F_{\textup{background,TESS2020}}$ & 0.0 -- 0.5      \\
	    $\sigma_{\textup{TESS}}$ & 0.0 -- 0.2  \\
	    \hline
		\multicolumn{2}{|c|}{\emph{Radial velocity parameters}}\\
		\hline
		$t_{0,\textup{rv}}$ (BJD-2 450 000) & 1523.04 -- 1523.06     \\
		$\gamma_{\textup{primary}} \textup{(km s$^{-1}$)}$ & 85 -- 95  \\
		$\gamma_{\textup{secondary}} \textup{(km s$^{-1}$)}$ &  65 -- 75    \\
		$\sigma_{\textup{primary}} \textup{(km s$^{-1}$)}$ & 0 -- 10 \\
		$\sigma_{\textup{secondary}} \textup{(km s$^{-1}$)}$ & 0 -- 10  \\
		\hline
	\end{tabular}
\end{table}

\begin{figure*}
	\includegraphics[width=1.8\columnwidth]{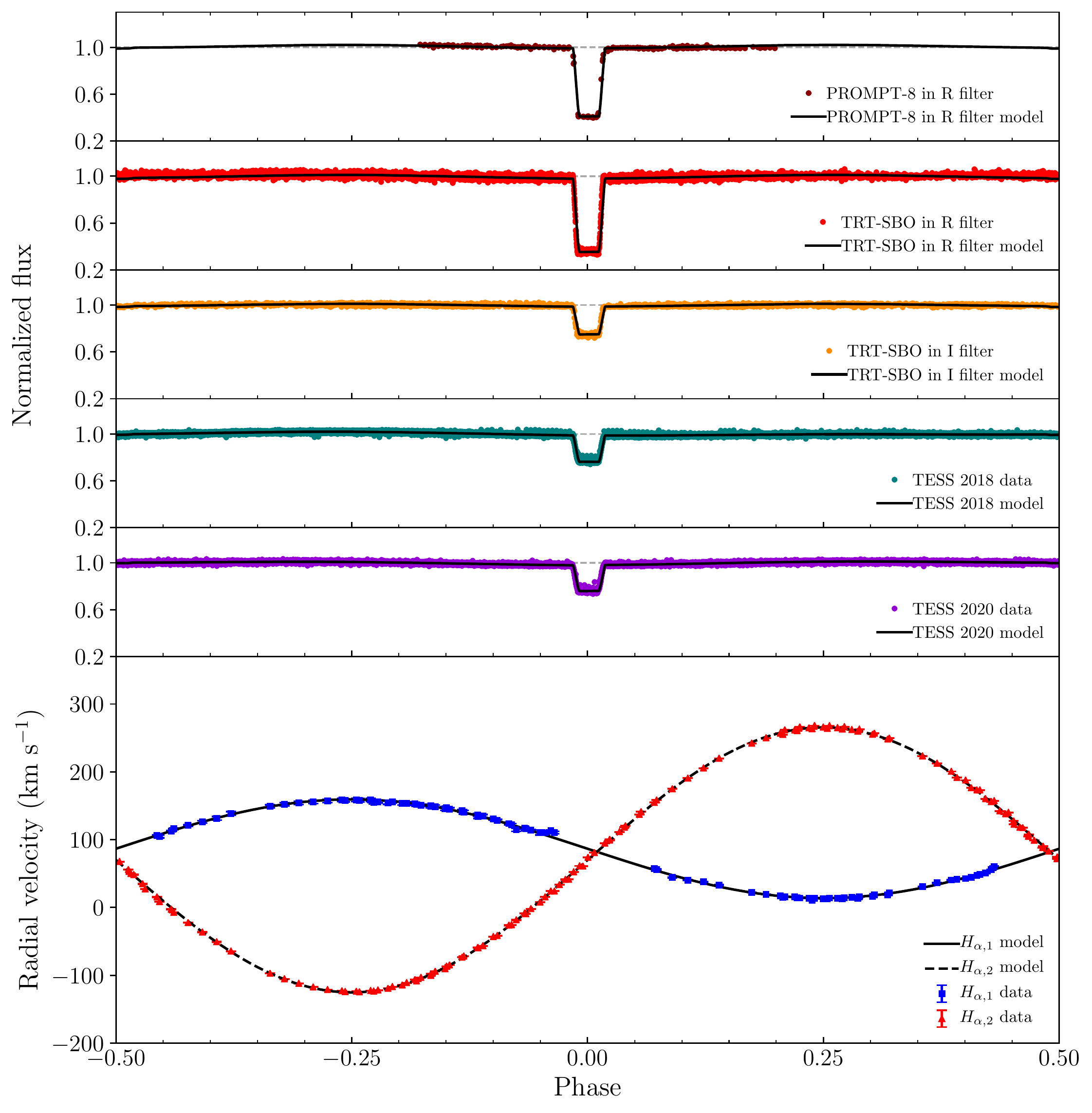}
    \caption{The RR Cae phase-folded light curves and H$_\alpha$ radial velocity curves with the best fitting models from the \texttt{PHOEBE} code. The light curves are obtained in the R-band from the PROMPT-8 telescope (Dark red), the R-band from the TRT-SBO (Red), the I-band from the TRT-SBO (Orange), the TESS in 2018 (Teal), and 2020 (Purple). The solid lines represent the fitted model from the \texttt{PHOEBE} code. The dashed lines show normalised baselines. The H$_\alpha$ radial velocity curves of primary and secondary components are obtained from the VLT data of Program ID: 076.D-0142 (PI: Maxted, Pierre) with the best fitting models of both the primary (Solid line) and secondary (Dashed-line) components from the \texttt{PHOEBE} code.}
    \label{fig:LC_RV}
\end{figure*}

\section{RR Cae physical parameters}
\label{sec:Parameters}
In order to obtain the physical parameters of RR Cae, the \texttt{PHOEBE} (PHysics Of Eclipsing BinariEs) version 2.3 \citep{Prsa2016,Conroy2020} was used to analyse the light curves. The \texttt{PHOEBE} code is based on the Wilson-Devinney code \citep{W-D1971}, which includes gravitational darkening, ellipsoidal variations, spots, etc. \citep{Prsa2016,Conroy2020}. The synthetic light curve from the \texttt{PHOEBE} code consists of the effects of gravitational darkening, ellipsoidal variations, reflection and heating, and starspots. The \texttt{emcee} \citep{emcee} python module that implements a Markov Chain Monte Carlo (MCMC) sampling algorithm was used with the \texttt{PHOEBE} to estimate the parameters' best-fit values. The log-likelihood function is given by

\begin{equation} \label{eq:lnlike}
\mathrm{ln} \, L (D|M) = -\frac{1}{2}\left[
\sum_{n}^{N_{LC}} \mathrm{ln}(2\pi\sigma^{2})+\sum_{n}^{N_{RV}} \mathrm{ln}(2\pi\sigma^{2})+ \chi^2\right ] \ .
\end{equation}
The $\chi^2$ function is calculated by 
\begin{dmath} \label{eq:lnlike_chi}
\chi^2 = \left[ \left ( \sum_{n}^{N_{LC}}\frac{(D_{\mathrm {obs,LC}}^{n} - M_{\mathrm{model,LC} }^{n} (\theta) )^2}{\sigma^{2}} \right ) + \left ( \sum_{n}^{N_{RV}}\frac{(D_{\mathrm {obs,RV}}^{n} - M_{\mathrm{model,RV} }^{n} (\theta) )^2}{\sigma^{2}} \right ) \right ] \ ,
\end{dmath}
where $D_{\mathrm {obs}}$ is the observed flux density, $M_{\mathrm{model}}$ is the modelled flux density and $\sigma^{2}$ is the variance of a flux measurement.

Due to the poor data quality from the PROMPT-8 and the TRT-SBO, only the TESS light curves are selected to fit along the radial velocity data from the VLT. In order to reduce the computation time, the light curves are phased-binned, with an orbital sampling period of 0.002 in the phase unit. We super-sampled the light curve because the observed light curves might be distorted in the long-cadence data, such as the TESS data \citep{Kipping2010}. Using the numerical estimation of \citep{Kipping2010}, the sampling of 0.002 in the phase unit of RR Cae is selected as it provides the maximum numerical integration error of $\sim$6,000 ppm, which is compatible with the photometric error of the TESS data (5,000-6,000 ppm).

The limb-darkening coefficients table of the TESS filters for the primary and the secondary are obtained from \cite{Claret2020DA} and \cite{Claret2018}, respectively. Each component's effective temperature ($T_{\rm eff}$) and surface gravity ($\log g$) are used to interpolate the limb-darkening coefficients. The reflective indices of both stars are set to be 0.6. The values are in the range of the reflective index of NN Ser (0.5-0.6), another WD+M4V binary \citep{Haefner2004}.

The TESS phased-binned light curve shows asymmetric maxima (Figure~\ref{fig:LC_TESS}). The out-of-eclipse variations in the light curve have been discussed in \citet{BruchandDiaz1998} and \citet{Bruch1999}. The variations might be caused by the ellipsoidal variations of a component in the system that has nearly filled its Roche lobe or by the effects of light reflection of one component to the other. Even though our simulated light curves using the \texttt{PHOEBE} code include these two effects, the simulated light curves do not show the asymmetric maxima. In order to create the asymmetric maxima light curves, we have to consider the other suggested effects indicating that the spot on the primary component might cause the asymmetric maxima light curves on RR Cae. Intriguingly, \citet{Ribeiro2013} suggested that a polar feature in the brightness distribution of the secondary component, similar to that observed in other fast-rotating stars, is the origin of the variation.

However, the asymmetric shape of the data taken in 2018 differs from the data in 2020, which occurred over a timescale of $\sim$2~yr. This change in the maxima of RR Cae light curves has yet to be reported in any previous study. In order to confirm the timescale of the evolution of the asymmetricity, the TESS data of every 20 consecutive orbits ($\sim$6~d) in both 2018 and 2020 are averaged and phase-binned with 0.005 phase resolution (Figure~\ref{fig:LC_TESS_Group}). For each year, the weekly average light curves show tiny variations. This result can be concluded that the light curves vary in longer timescales, such as monthly or yearly timescales.

In order to explain the origin of these variations, the model with a starspot is adopted, as suggested by \citet{Maxted2007}. In this work, we focus on the model with only a starspot. Although the two-spot model might be a better explanation for the change in the asymmetric shape of the data, the two-spot model is excluded from the scope of this work due to the complexity of the two-spot model, such as the overlap of the spots. Only a spot variation model is focused on in this work. The two models of spot variations corresponding to two different years (2018 and 2020) on the secondary component are used. A spot model consists of four parameters: the spot temperature in a unit of stellar temperature ($T_{\textup{rel,eff}}$), the spot radius ($\psi$), the spot co-latitude ($\beta$), and the spot longitude ($\lambda$).

Moreover, the TESS data in 2018 and 2020 show differences in the primary eclipse depths. The primary eclipse depth of TESS data in 2018 is 23.7$\%$, while the depth in 2020 is 24.1$\%$. The depth of the primary eclipse in 2018 was shallower than the primary eclipse depth in 2020. The difference might be caused by the large TESS pixel size. As the large TESS pixel size, the flux background source might be blended in the RR Cae's photometric pixels. In order to reduce the blending effect, the model with a blending background flux is adopted:

\begin{equation} 
\label{eq:f_background}
F_{\textup{normalised}} = F_{\textup{binary}} + F_{\textup{background}} \ ,
\end{equation}
where $F_{\textup{normalised}}$ is a normalised observed flux. $F_{\textup{binary}}$ is a flux from a binary system modelled by \texttt{PHOEBE} code, and $F_{\textup{background}}$ is a blending factor in the unit of the normalised observed flux.

As the system showed a significant eclipse timing variation, as shown by \citeauthor{Qian2012}, the time of minima in the TESS data observed in 2018 and 2020 differ from the time of minima in 2005, when the spectra are obtained from the VLT by a few tenths seconds. Therefore, we separately reported the reference times of minimum, $t_0$, obtained from the TESS light curves fitting, $t_{0,\textup{TESS}}$, and the VLT radial velocity fitting, $t_{0,\textup{rv}}$.

\onecolumn
\begin{landscape}
\begin{table*}
	\begin{center}
	\caption{The physical parameters of RR Cae.}
	\label{tab:physical_model}
	\begin{tabular}{lcccccc}
		\hline \hline
		\multirow{2}{*}{Parameters} &
		\multicolumn{2}{c}{Fitted values} & \multicolumn{4}{c}{Literature values} \\
		\cline{2-3} \cline{4-7} 
        & Model without spot & Model with spot & \citet{Ribeiro2013} & \citet{Maxted2007} & \citet{Bruch1999} & \citet{BruchandDiaz1998} \\
		\hline
		\multicolumn{7}{|c|}{\emph{System parameters}}\\
		\hline
		$t_{0}$ (BJD-2 450 000)  & $1523.052^{+0.006}_{-0.006}$  & $1523.0493^{+0.0003}_{-0.0004}$ & -- & 1523.048 567 $\pm$ 0.000 019   & -- &  (HJD)2 450 685.433 42 $\pm$ 0.000 13 \\
		$P_{0}$ (d) &  0.303 703 5$^{+2\textup{E}-7}_{-3\textup{E}-7}$ & 0.303 703 61$^{+1\textup{E}-8}_{-2\textup{E}-8}$ & -- & 0.303 703 6366 $\pm$ 0.000 000 0047 & -- & 0.303 695 $\pm$ 0.000 014 \\
		$i$ (degree)   & $82.4^{+4.9}_{-0.4}$  & $82.9^{+0.2}_{-0.3}$ & 79 & 76 -- 90 & 84.04 $\pm$ 0.17 & 87.9   \\
		$a$ ${(R_\odot)}$ &  $1.63^{+0.01}_{-0.03}$ & $1.622^{+0.003}_{-0.003}$ & -- & 1.370 $\pm$ 0.007 & 1.37 $\pm$ 0.07 & 1.46     \\
		$T_{1}$ (K) & 7260$^\star~^\dagger$ & 7260$^\star~^\dagger$ & 7260 $\pm$ 250  & 7540 $\pm$ 175 & 7005 $\pm$ 85 & 7005 \\
		$T_{2}$ (K) & $2930^{+140}_{-40}$ & $2730^{+90}_{-60}$ & 3500$^\dagger$ & 3100 $\pm$ 100 & 2500 $\pm$ 15 & 2765 \\
		$q$         & $0.371^{+0.003}_{-0.002}$ & $0.371^{+0.002}_{-0.002}$ & 0.376 $\pm$ 0.0005  & 0.413 -- 0.416 & 0.204 $\pm$ 0.005 & 0.243                 \\
		$M_{1}$ ${(\textup{M}_\odot)}$  & 0.46 $\pm$ 0.12 & 0.453 $\pm$ 0.002 & 0.43 $\pm$ 0.02 & 0.440 $\pm$ 0.023 & 0.467  & 0.394  \\
		$M_{2}$ ${(\textup{M}_\odot)}$  & 0.17 $\pm$ 0.07 & $0.168 \pm 0.001$ & 0.15  & (0.182 -- 0.183) $\pm 0.012$ & 0.095  & 0.089      \\
		$R_{1}$ ${(R_\odot)}$  & $0.03^{+0.01}_{-0.03}$ & $0.020^{+0.001}_{-0.001}$ & -- & $(0.015 -- 0.016)~\pm 0.0004$ & 0.0152 & 0.0162     \\
		$R_{2}$ ${(R_\odot)}$  & $0.26^{+0.01}_{-0.01}$ & $0.247^{+0.007}_{-0.006}$ & 0.21 & (0.203 -- 0.215) $\pm 0.015$ & 0.189 & 0.134 \\
	    \hline
		\multicolumn{7}{|c|}{\emph{Light curve parameters}}\\
		\hline
		$F_{\textup{background,TESS2018}}$ & $0.007^{+0.005}_{-0.110}$  & $0.02^{+0.04}_{-0.02}$ & -- & -- & -- & -- \\
		$F_{\textup{background,TESS2020}}$ & $0.002^{+0.001}_{-0.035}$ & $0.09^{+0.04}_{-0.03}$ & -- & -- & -- & -- \\
		$\sigma_{\textup{TESS}}$ & 
		$2\times10^{-5}$ $ ^{+1\textup{E}-5}_{-3\textup{E}-5}$ & $1\times10^{-7}$ $ ^{+1\textup{E}-7}_{-1\textup{E}-7}$ & -- & -- & -- & -- \\
	    \hline
		\multicolumn{7}{|c|}{\emph{Radial velocity parameters}}\\
		\hline
		$t_{0,\textup{rv}}$ (BJD-2 450 000) & $1523.050^{+0.002}_{-0.002}$  & $1523.0486^{+0.0001}_{-0.0001}$ & -- & -- & -- & --  \\
		$\gamma_{\textup{primary}} \textup{(km s$^{-1}$)}$ & $86.7^{+0.3}_{-0.3}$ & $87.0^{+0.2}_{-0.3}$ & $95.1~\pm~0.3$ & $81.5~\pm~2.1$ & -- & --   \\
		$\gamma_{\textup{secondary}} \textup{(km s$^{-1}$)}$  & $70.1^{+0.4}_{-0.2}$ & $70.1^{+0.2}_{-0.2}$ & $77.9~\pm~0.4$ & $85.8~\pm~3.6$ & $40~\pm~5$ & --  \\
		$\sigma_{\textup{primary}} \textup{(km s$^{-1}$)}$ & $4.0^{+0.8}_{-1.4}$  & $4.3^{+1.2}_{-0.8}$ & -- & -- & -- & -- \\
		$\sigma_{\textup{secondary}} \textup{(km s$^{-1}$)}$  & $4.2^{+0.6}_{-1.3}$ & $3.8^{+0.6}_{-0.4}$ & -- & -- & -- & -- \\
		\hline
	\end{tabular}
	\end{center}
	\begin{flushleft}
	Remarks: $^\star$: The primary component temperature is obtained from \citet{Ribeiro2013}. $^\dagger$: Fixed value.
	\end{flushleft}
\end{table*}
\end{landscape}
\twocolumn

\begin{table}
	\centering
	\caption{The parameters of a spot on the secondary component of RR Cae in 2018 and 2020.}
	\label{tab:spot_model}
	\begin{tabular}{lccc}
		\hline \hline
		\multirow{2}{*}{Parameters} & \multirow{2}{*}{Prior} & \multicolumn{2}{c}{Fitted values}  \\
		\cline{3-4}
		& & Year 2018 & Year 2020 \\
		\hline
		$T_{\textup{rel,eff}}$  & 0.2 -- 2.0 & $1.021^{+0.0012}_{-0.0006}$ & $0.993^{+0.003}_{-0.002}$  \\
		$\psi$ (degree) & 1 -- 100 & $82^{+10}_{-13}$ & $57^{+11}_{-8}$    \\
		$\beta$ (degree) & 0 -- 180 & $14^{+6}_{-4}$ &  $54^{+20}_{-10}$  \\
	   	$\lambda$ (degree) & 0 -- 360 & $287^{+3}_{-3}$ & $190^{+2}_{-2}$   \\
		\hline
	\end{tabular}
\end{table}

\begin{figure*}
	\includegraphics[width=2\columnwidth]{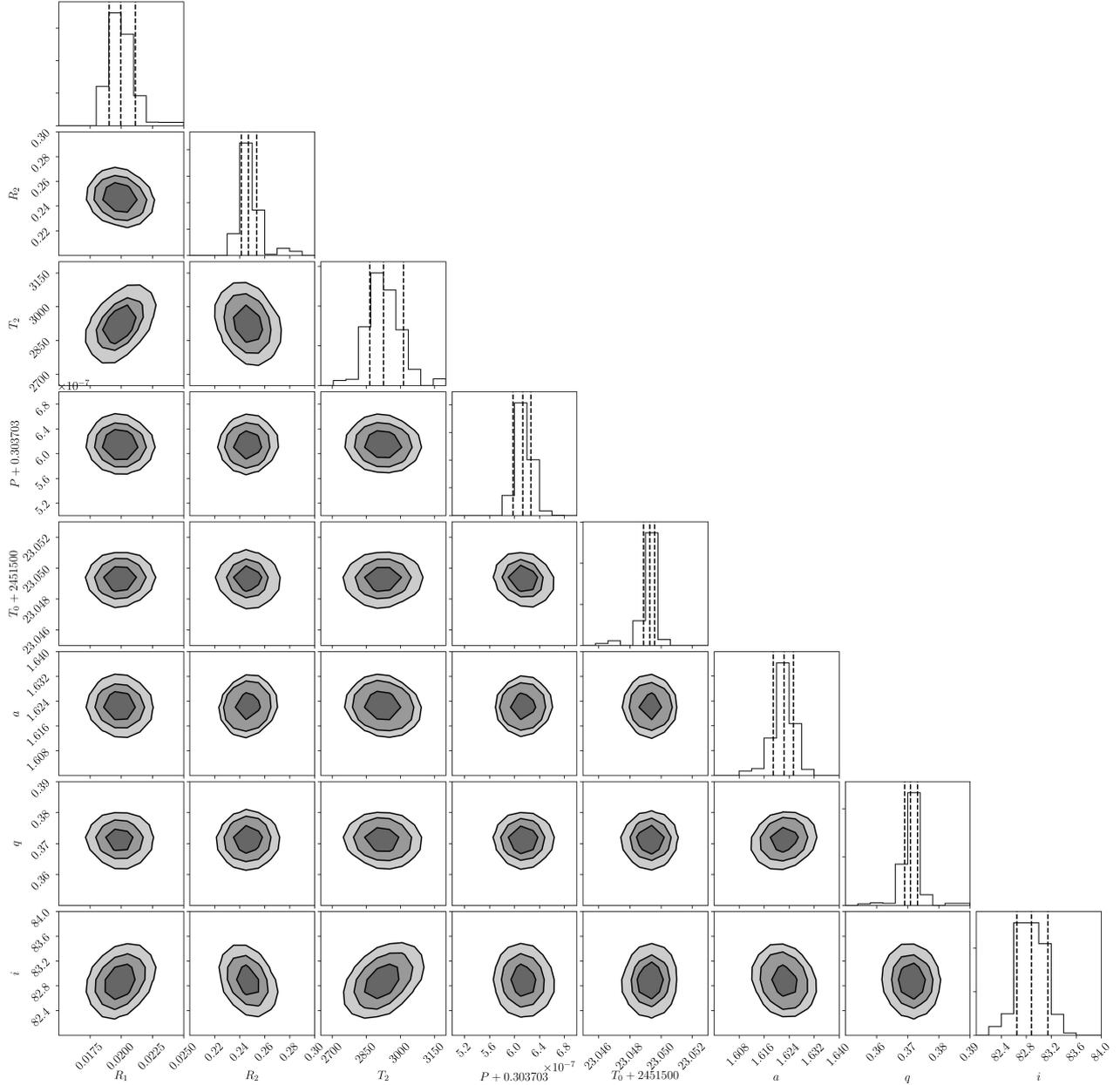}
    \caption{Posterior probability distributions of the physical parameters of RR Cae using the \texttt{PHOEBE} code with the MCMC fitting. The dashed-lines mark the 16th, 50th, and 84th percentiles.}
    \label{fig:param_physical}
\end{figure*}

\begin{figure}
	\includegraphics[width=1\columnwidth]{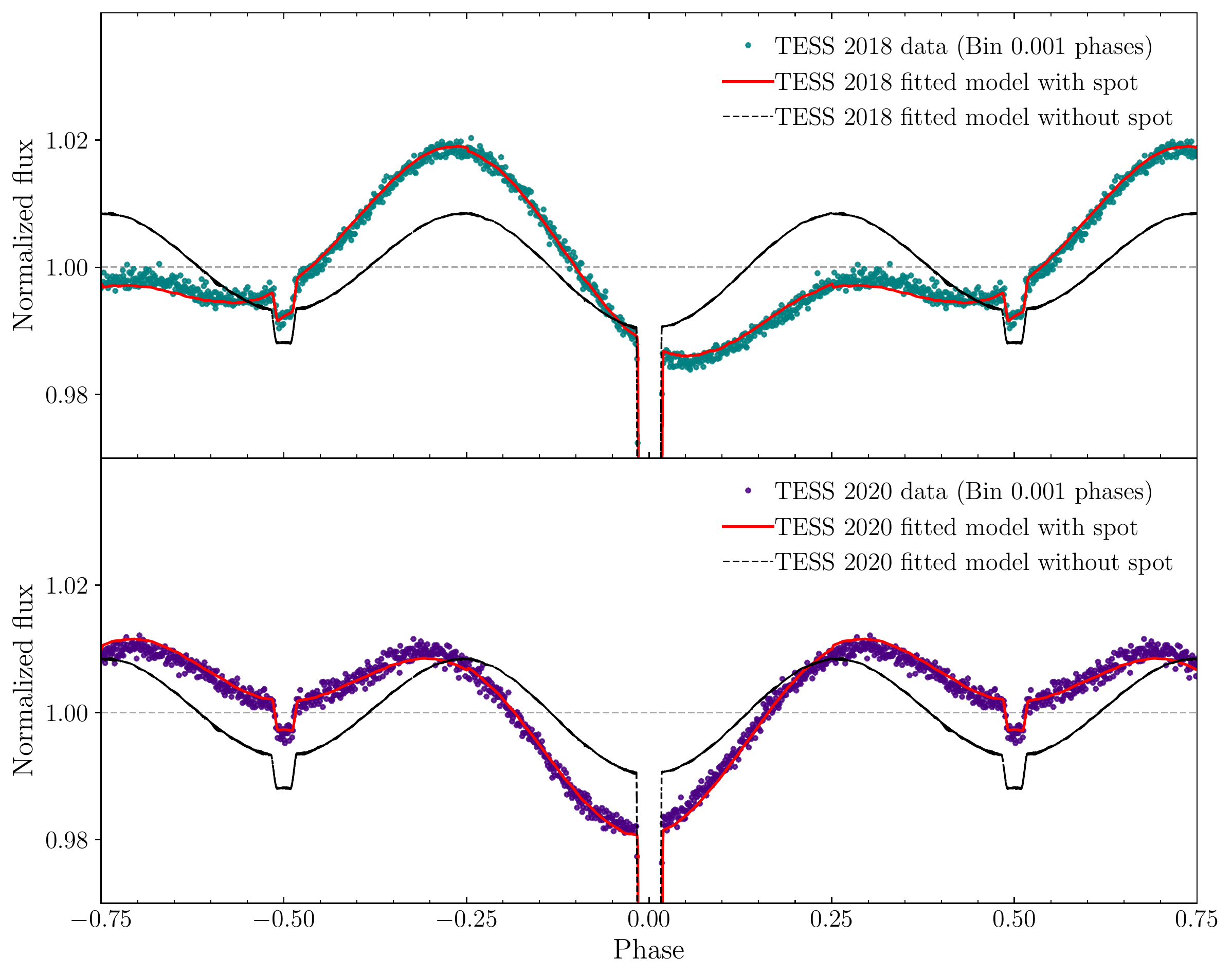}
    \caption{The RR Cae phase-folded light curves obtained from the TESS in 2018 (Teal) and 2020 (Purple). The red solid lines represent the best fitting model with a starspot on the secondary component and the black dashed lines represent the best fitting model without starspot from the \texttt{PHOEBE} code.}
    \label{fig:LC_TESS_model}
\end{figure}

\begin{figure}
	\includegraphics[width=\columnwidth]{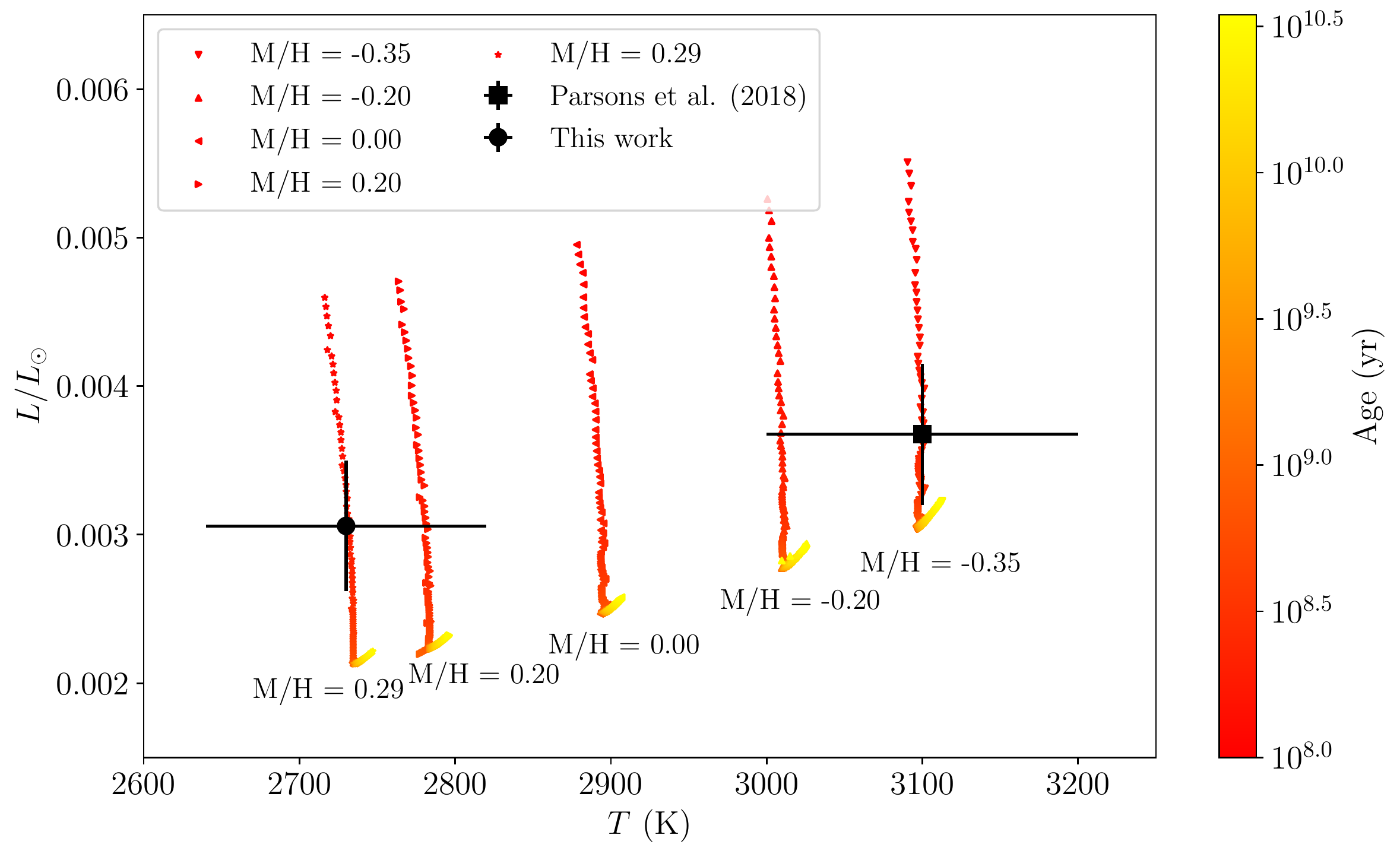}
    \caption{The isochrones of a star with the mass of $0.168~\textup{M}_\odot$. The black square represents the dwarf with parameters from \citet{Parsons2018} and the black circle represents the dwarf with parameters from the \texttt{PHOEBE} best fitting model.}
    \label{fig:iso}
\end{figure}

\begin{figure*}
	\includegraphics[width=2\columnwidth]{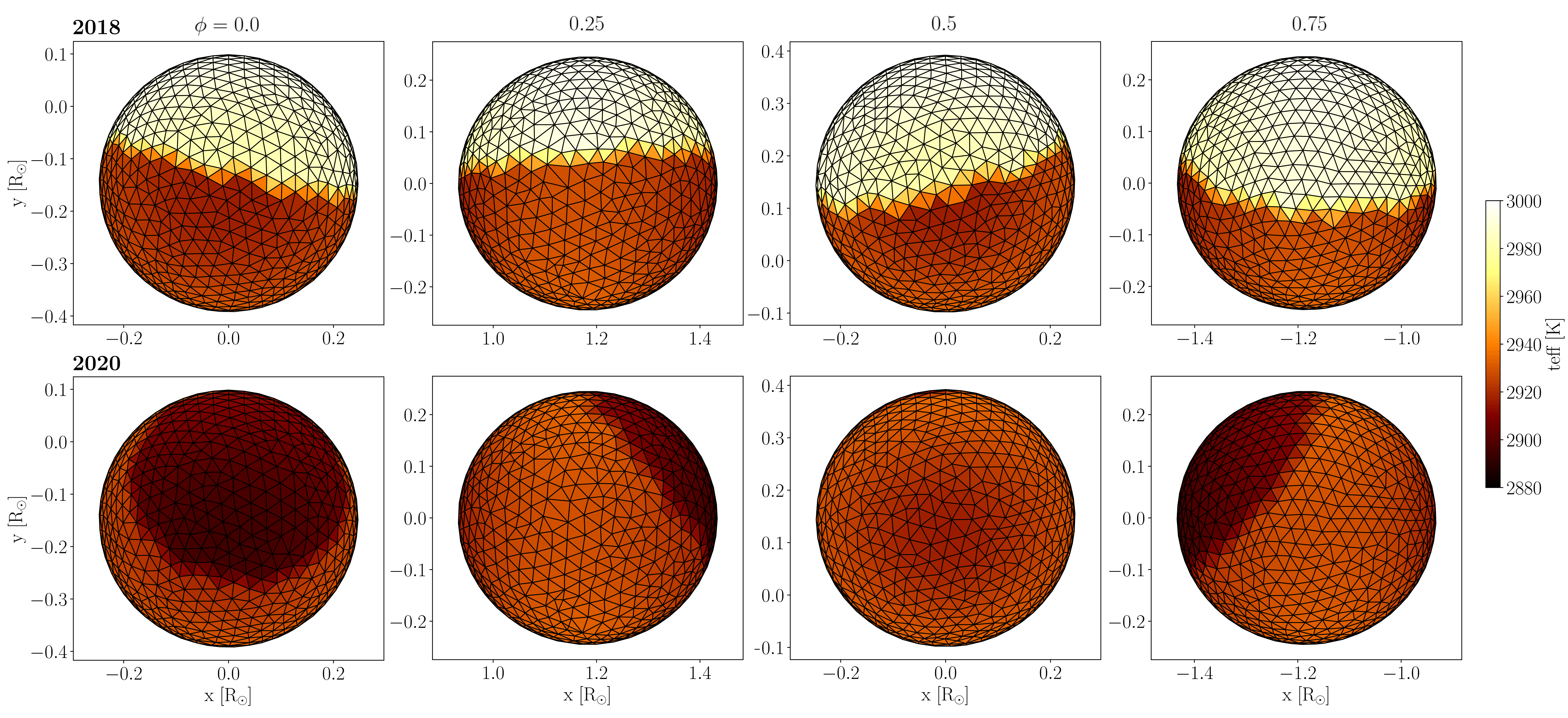}
    \caption{The spot models on the secondary component of RR Cae at the different orbital phases: 0.00, 0.25, 0.50, 0.75 (Left to Right) in 2018 (Top) and 2020 (Bottom).}
    \label{fig:spot_model}
\end{figure*}

For the MCMC procedure, the fitting parameters and their priors are shown in Table~\ref{tab:prior}. The uniform priors are adopted. The number of walkers is ten times the number of fitting parameters. The burn-in and production chains are 2000 and 3000 steps, respectively. The primary component temperature is fixed as the obtained temperature from \citet{Ribeiro2013}. The fitted light curves and radial velocity curves are shown in Figure~\ref{fig:LC_RV} The physical parameters results for the MCMC analyses of the model with starspot are shown in Figure~\ref{fig:param_physical}. In order to confirm that the spot is needed in the modelling, the light curves are also fitted with the model without a spot (Table~\ref{tab:physical_model} and Figure~\ref{fig:LC_TESS_model}). The result shows that the \texttt{PHOEBE} model, which already includes the effects of the tidal variations and the reflection, cannot fit RR Cae's out-of-eclipse variations. Therefore, the model with a spot is used for the fitting in this work.

The physical parameters of RR Cae from \texttt{PHOEBE} and previous literature are shown in Table~\ref{tab:physical_model}. \citet{BruchandDiaz1998} and \citet{Bruch1999} suggested that the system has a low mass ratio ($q < 0.3$) with high orbital inclination ($i > 80^\circ$), while \citet{Maxted2007} and \citet{Ribeiro2013} works which include radial velocity data prefer the models with higher mass ratio ($q > 0.35$) and lower orbital inclination ($i < 80^\circ$). The fitting values from \texttt{PHOEBE} provide a system orbital inclination of $82.9^{+0.2}_{-0.3}$ deg with a mass ratio of $0.371^{+0.002}_{-0.002}$, which is corresponding to the primary component mass, $M_1$, of $0.453^{+0.002}_{-0.002}~\textup{M}_\odot$ and the secondary component mass, $M_2$, of $0.168^{+0.001}_{-0.001}~\textup{M}_\odot$. The fitted masses agree with the masses from \citet{Ribeiro2013}. The binary orbital separation, $a$, of $1.622^{+0.003}_{-0.003}$ is obtained from \texttt{PHOEBE}. The value is larger than the separations from the other works. The dwarf effective temperature ($2730^{+90}_{-60}$ K) is cooler than the temperature obtained from \citet{Maxted2007} and \citet{Ribeiro2013}. 

As a very precise mass of the secondary component is obtained from the fitting, the theoretical isochrones are used to test its mass-radius relationship. The isochrones are obtained from the \texttt{CMD} version 3.6\footnote{\texttt{CMD} v3.6: \texttt{http://stev.oapd.inaf.it/cgi-bin/cmd_3.6}}. The \texttt{CMD} version 3.6 code is based on PARSEC code version 1.25, which includes the model of very low mass stars in the code \citep{Bressan2012,Chen2014}. The isochrones of a star with a mass of $0.168~\textup{M}_\odot$ are shown in Figure~\ref{fig:iso}. \citet{Parsons2018} reported that the RR Cae dwarf component has [Fe/H]$=-0.35$ with the age of $6.11^{+0.65}_{-0.47}$ Gyr. \citet{Parsons2018} dwarf's parameters still favour the \texttt{CMD} isochrone with [M/H]$=-0.35$. However, the \texttt{CMD} isochrone provides the age of the dwarf $\sim$0.2~Gyr, which is 30 times younger than the age reported by \citet{Parsons2018}. When comparing the parameters obtained from \texttt{PHOEBE} in this work with the isochrones from the \texttt{CMD} code, the parameters fit the isochrones with metallicities [M/H]$=0.29$ and the age of $\sim$0.2~Gyr.

For the fitting model with starspots, the TESS photometric data provides the linear ephemeris of
                
\begin{equation}
\label{eq:ephermeris_TESS}
    \textup{BJD}=2\ 451\ 523.049\ 3 (3)+0.303\ 703\ 6(1).E \ .
\end{equation}
However, the linear ephemeris of the UVES radial velocity data is
        
\begin{equation}
    \textup{BJD}=2\ 451\ 523.048\ 6 (1)+0.303\ 703\ 6(1).E \ .
\end{equation}
From the ephemerides, the TESS reference time of minimum ($t_{0,\textup{TESS}}$) differs from the reference time of minimum of radial velocity data from the VLT ($t_{0,\textup{rv}}$) by 0.0007 d or 1 min. The difference in the reference time of minima is caused by the timing variation of the system, which will be discussed in Section~\ref{sec:OC}. As the TESS light curves show the out-of-eclipse variation and spots are added on the secondary component, the spot parameters are shown in Table~\ref{tab:spot_model}. The spot illustrations are shown in Figure~\ref{fig:spot_model}. In 2018, a hot spot with a radius of 80$^\circ$ is found near the north pole of the dwarf ($\beta_{2018}~=~14^{+6}_{-4}$ deg). While in 2020, a smaller cold spot is found at moderate latitude ($\beta_{2020}~=~54^{+20}_{-10}$ deg). Note that the relative temperature between hot and cold spots, and the star is less than $3\%$, which is smaller than the uncertainty of the effective temperature of the secondary component obtained from the fitting model ($2730^{+90}_{-60}$~K).

It is difficult to explain such an evolution of large spots on the dwarf over a time span of two years. However, the heat transfer between two components might explain such an evolution as discussed in Section~\ref{sec:Spectrum}. The heated surface could act as a large spot on the surface of the secondary component.

\section{RR Cae time of minima and its circumbinary planets}
\label{sec:OC}

\subsection{RR Cae eclipse time variation}
\label{sec:ETV}

In Sections~\ref{sec:Observation}, 20, 4, and 406 epoch light curves from the TESS, the PROMPT-8 telescope, and the TRT-SBO are provided, respectively. Although the linear ephemeris of the TESS photometric data is already obtained from the parameters fitting in the previous section, the fitting might not provide precisely fitted timings as it used the phased-binned light curves for the fitting. In order to obtain the precise time of the minimum of each primary eclipse, we refit each eclipse using the physical parameters obtained from the phased-binned fitting (Table~\ref{tab:physical_model}). Moreover, we increased the sampling rate of the synthetic light curves with \texttt{PHOEBE} to 0.1~s time resolution for the fitting. The synthetic light curves are simulated in three filters: the R-, the I- and the TESS bands, with \texttt{PHOEBE}. The light curves in the R- and the I-bands are simulated with the interpolated limb-darkening coefficients from \citet{Claret2020} and \citet{Claret1998} for primary and secondary stars, respectively. Due to spot variation and the poor quality of out-of-eclipse data in the R- and the I- bands, the R- and the I-bands light curves have not added a spot in the secondary component, and only the data with phases between -0.1 and 0.1 are used. The R-band data from the PROMPT-8 telescope and the TRT-SBO show different eclipse depths. Therefore, the difference $F_{\textup{Background}}$ is adopted. The simulated light curves are interpolated with the spline interpolation and performed the MCMC fitting to find the minimum time of each eclipse. The fitted times of minima are shown in Table~\ref{tab:ToM}.

\begin{figure}
	\includegraphics[width=1\columnwidth]{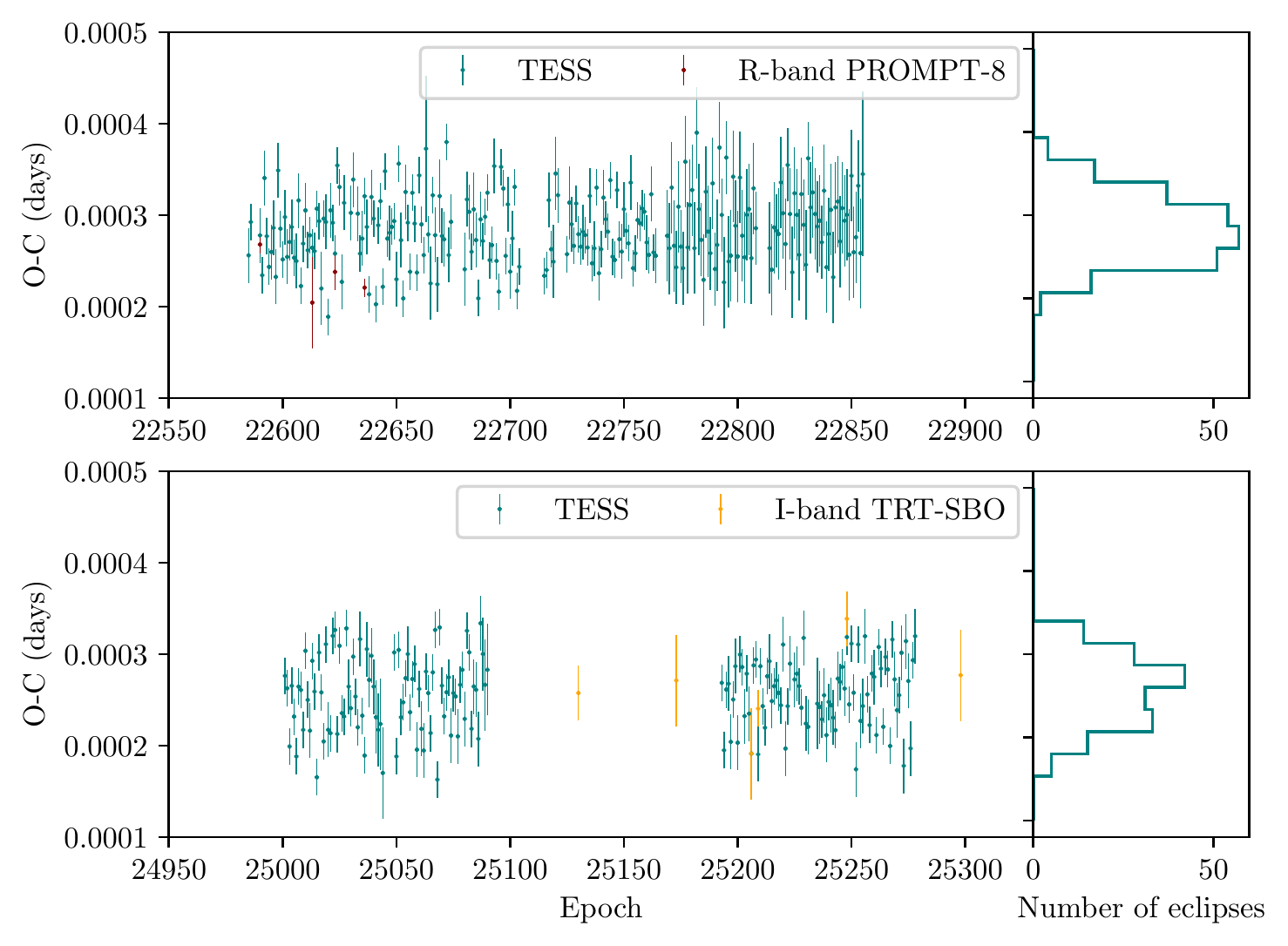}
    \caption{The RR Cae O-C diagrams (Left) in 2018 (Top) and 2020 (Bottom) were obtained in this work with the distributions of the TESS observations (Right). The distributions of the TESS observations can be seen as a Gaussian distribution.}
    \label{fig:oc_hist}
\end{figure}

\begin{figure}
	\includegraphics[width=1.\columnwidth]{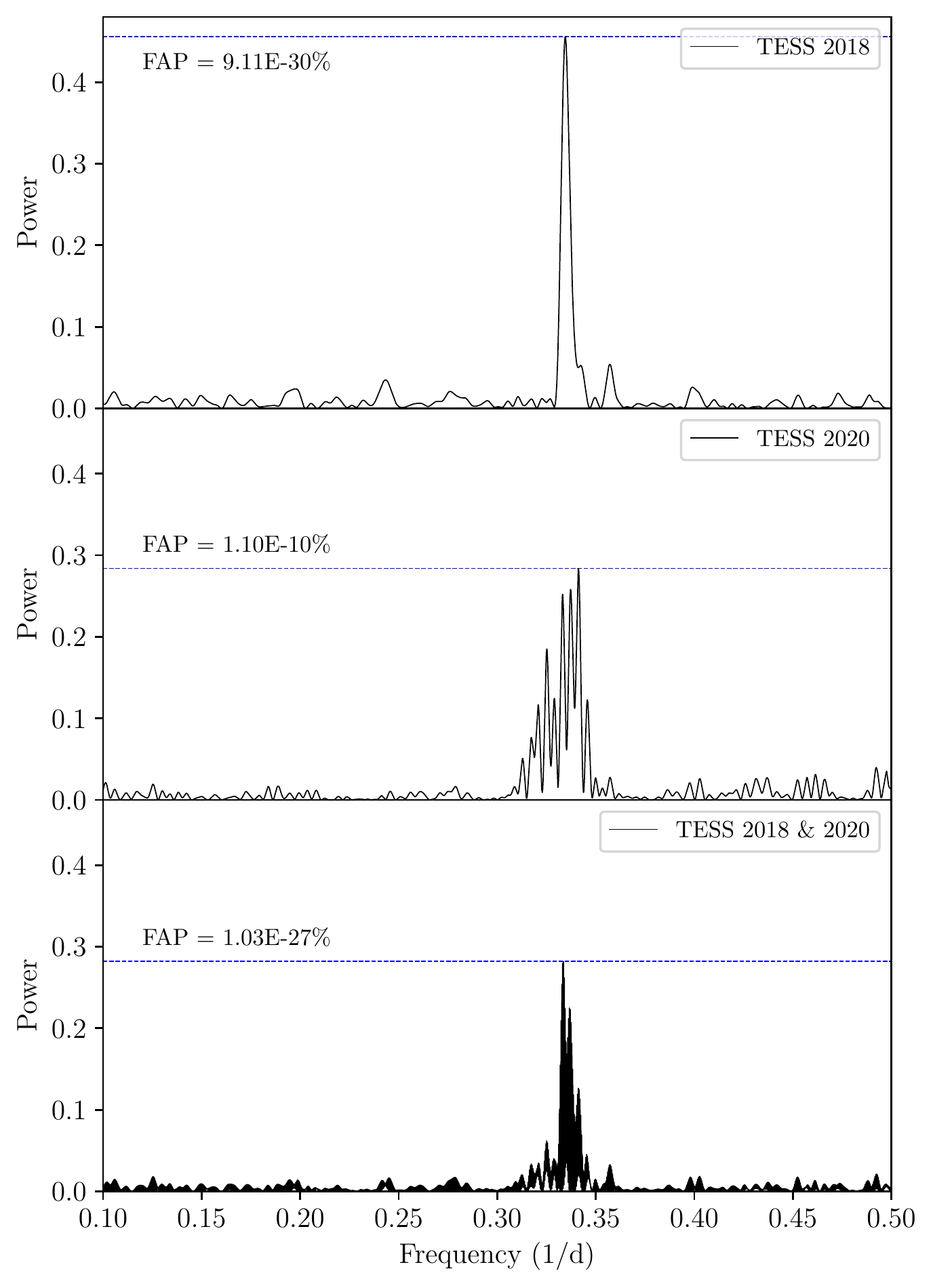}
    \caption{The periodograms of the O-C diagram using the TESS observations obtained in 2018 (Top), 2020 (Middle), and combined both years (Bottom). The periodogram of the combined TESS data in 2018 and 2020 shows the highest amplitude at the frequency of 0.3336 $\textup{cycles}^{-1}$ with the FAP of $1.03\times 10^{-27}\%$.}
    \label{fig:oc_lomb}
\end{figure}

\begin{figure}
	\includegraphics[width=1.05\columnwidth]{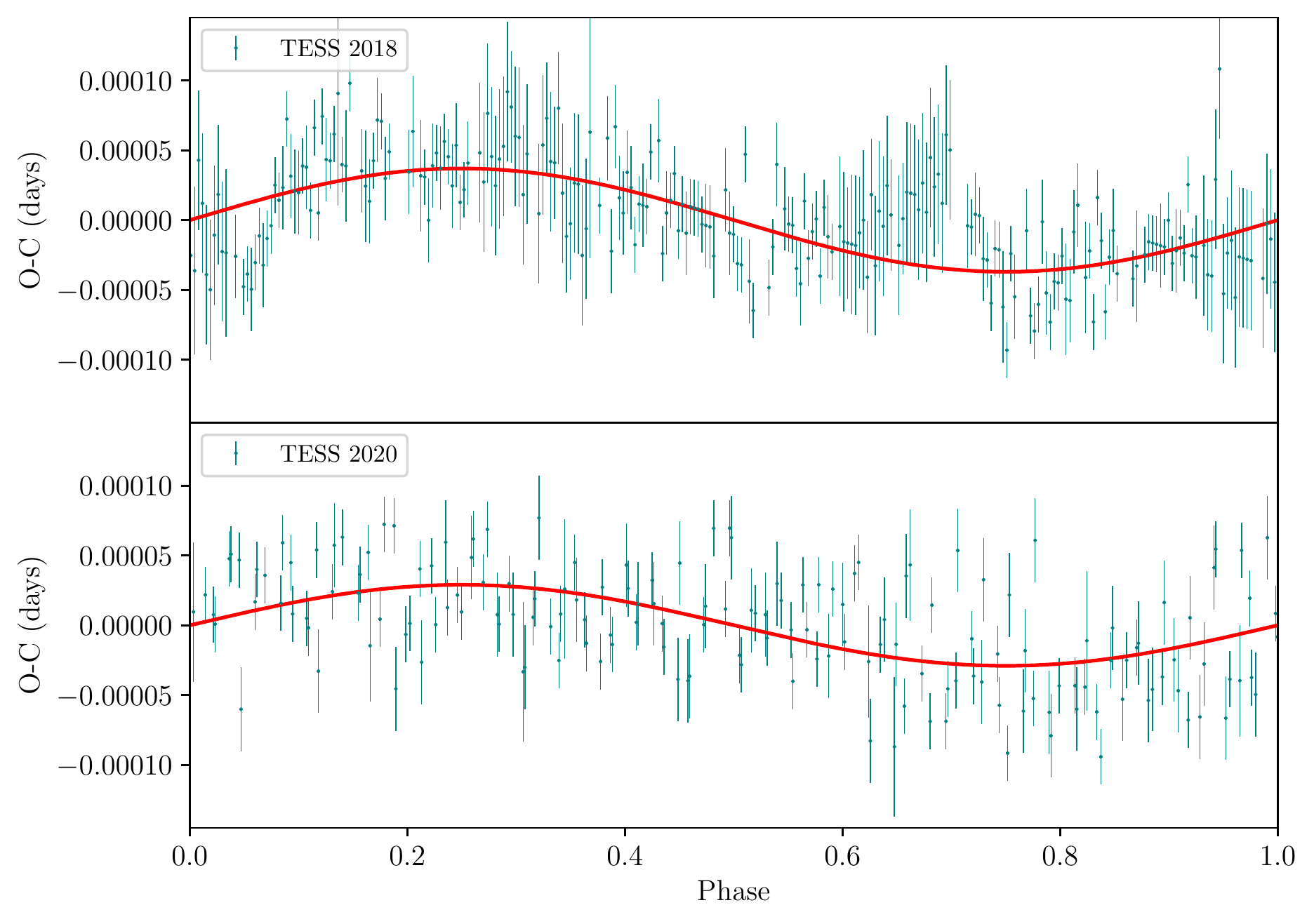}
    \caption{The phase-folded O-C diagrams from the TESS data in 2018 (Top) and 2020 (Bottom). The O-C diagrams are phase-folded by the highest amplitude frequency in Figure~\ref{fig:oc_lomb} (0.3336 cycle$^{1}$). The red lines show the sinusoidal function with the frequency.}
    \label{fig:oc_phase}
\end{figure}

Combining the eclipse time from the TESS, the PROMPT-8 telescope, and the TRT-SBO, the eclipse times from the TRT-SBO and the PROMPT-8 telescope are consistent with the timing from the TESS in 2018 and 2020 (Figure~\ref{fig:oc_hist}). In order to analyse the short-time scale cyclic variation eclipse timing, the Generalized Lomb-Scargle Periodogram of \texttt{PyAstronomy} is applied to the TESS data \citep{pya}. The TESS data in both 2018 and 2020 show that the highest powers are at periods of 2.989 and 2.930 cycles with the false alarm probabilities of $9.11\times10^{-30}\%$ and $1.10\times10^{-10}\%$, respectively (Figure~\ref{fig:oc_lomb}). Combining the data in 2018 and 2020 has the highest power period at 2.997 cycles with false alarm probabilities of $1.03\times10^{-27}\%$. In Figure~\ref{fig:oc_phase}, the best sine fits of the data in 2018 and 2020 are shown. These short cyclic variations might be caused by the stellar activity or the starspot \citep{Watson2004,Conroy2014}. As the cyclic variations have a period $\sim$3 epoch, the TESS light curves are grouped by remainder integers when we divide the RR Cae's epochs by three to investigate the $\sim$3 epoch cyclic variations. There is no significant difference between these three datasets. Moreover, these light curves do not show any significant cyclic variations from the stellar activity or the starspots. As the amplitude of this cyclic eclipse time variation (Figure~\ref{fig:oc_phase}) is smaller than the TESS cadence, continuous high-time precision monitoring on RR Cae is needed to confirm these short variations in the future.

\begin{table}
	\caption{The lists of time of minima of RR Cae's primary eclipse.}
	\label{tab:ToM}
	\begin{tabular}{lcllc}
		\hline
		Observatory & Filter & $t_0$ (+2 400 000 BJD) & Error (d) & Ref\\
		\hline
		LCO  & B & 45 927.916 650 & 0.000 116  & (1) \\
		SAAO & W & 49 721.478 524 & 0.000 003 & (2)  \\
		SAAO & W & 49 722.389 675 & 0.000 003 & (2) \\
		SAAO & W & 49 726.337 828 & 0.000 004 & (2) \\
        LNA & R & 50 681.789 580 & 0.000 116 & (3) \\
		LNA & R & 50 684.827 030 & 0.000 116 & (3) \\
		LNA & R & 50 687.863 610 & 0.000 116 & (3) \\
		LNA & R & 50 688.774 700 & 0.000 116 & (3) \\
		LNA & R & 50 688.774 770 & 0.000 116 & (3) \\
		SAAO & W & 50 700.619 201 & 0.000 004 & (2) \\
		SAAO & W & 50 750.426 547 & 0.000 002 & (2)\\
		SAAO & W & 50 753.463 587 & 0.000 002 & (2) \\
		SAAO & W & 50 756.500 622 & 0.000 002 & (2) \\
		SAAO & W & 51 045.626 463 & 0.000 002 & (2) \\
		SAAO & I & 51 523.352 26 & 0.000 03 & (2)\\
		SAAO & W & 51 524.567 06 & 0.000 05 & (2) \\
		SAAO & I & 51 524.567 07 & 0.000 03 & (2) \\
		SAAO & I & 51 528.515 18 & 0.000 04 & (2) \\
		SAAO & I & 51 532.463 35 & 0.000 03 & (2) \\
		SAAO & W & 53 228.648 145 & 0.000 002 & (2) \\
		VLT & $u^{'}$ & 53 701.514 8245 & 0.000 0019 &  (4) \\
		VLT & $g^{'}$ & 53 701.514 8236 & 0.000 0008 & (4) \\  
		VLT & $i^{'}$ & 53 701.514 8346 & 0.000 0034 & (4) \\
		VLT & $u^{'}$ & 53 701.818 5392 & 0.0000 032 & (4) \\
		VLT & $g^{'}$ & 53 701.818 5359 & 0.000 0004 & (4) \\  
		VLT & $i^{'}$ & 53 701.818 5315 & 0.000 0019 & (4) \\
		CASLEO & V & 55 533.759 31 & 0.000 05 & (5) \\
		CASLEO & V & 55 537.707 49 & 0.000 05 & (5) \\
		CASLEO & N & 55 655.544 47 & 0.000 02 & (5) \\
		CASLEO & N & 55 889.699 91 & 0.000 02 & (5) \\
		CASLEO & N & 55 891.825 85 & 0.000 02 & (5) \\
		CASLEO & N & 55 892.736 95 & 0.000 02 & (5) \\
		TESS & TESS & 58 382.195 39 & 0.000 03 & (6) \\
		TESS & TESS & 58 382.499 13 & 0.000 02 & (6) \\
		PROMPT-8 & R & 58 383.713 92 & 0.000 01 & (6) \\	
		TESS & TESS & 58 383.713 93 & 0.000 03 & (6) \\
		TESS & TESS & 58 384.017 59 & 0.000 02 & (6) \\
		TESS & TESS & 58 384.321 40 & 0.000 03 & (6) \\
		$\dots$ & $\dots$ & $\dots$ & $\dots$ & $\dots$ \\
		$\dots$ & $\dots$ & $\dots$ & $\dots$ & $\dots$ \\
		$\dots$ & $\dots$ & $\dots$ & $\dots$ & $\dots$ \\
		\hline
	\end{tabular}
\textbf{Note.} \\
Observatory: LCO: Las Campanas Observatory, Cerro Las Campanas, Chile. SAAO: South African Astronomical Observatory, Sutherland, South Africa. LNA: Laboratorio Nacional de Astrofisica, Pico dos Dias, Brazil. VLT: Very Large Telescope. CASLEO: Jorge Sahade telescope at Complejo Astronomico El Leoncito, San Juan, Argentina. \\
\textbf{References:} \\ (1) \citet{Krzeminski1984}, (2) \citet{Maxted2007}, (3) \citet{BruchandDiaz1998}, (4) \citet{Parsons2010}, (5) \citet{Qian2012} and (6) This study
\end{table}

\subsection{Light travel time effect}
\label{sec:LTT}

\citet{Qian2012} showed that the eclipse timing variation of RR Cae might be caused by the gravitational interaction of a circumbinary planet with a planet mass of 4.2~M$_{\textup{Jup}}$ and an orbital period of 11.9 yr. In order to investigate the existence of the planet, the obtained times of minima in Section~\ref{sec:ETV} are combined with primary eclipse times from \citet{Parsons2010} and \citet{Qian2012}, as shown in Table~\ref{tab:ToM}. The times of minima are fitted with the binary period change model combined with the light travel time effect from a circumbinary object as follows,

\begin{equation}
    t = t_0 + P \times E + \frac{1}{2}\frac{\textup{d}P}{\textup{d}E} E^2 + \sum_{n=3}^{N} \tau{_n},
	\label{eq:oc_2LTT}
\end{equation}
where $t$ is the time of minimum, $t_0$ is the reference time of minimum, $P$ is the orbital period of the binary system, $E$ is the observed epoch and $\tau$ is the sinusoidal term for light travel time (LTT) effect by the third component \citep{Irwin1952},

\begin{equation}
    \tau{_n} = K{_n} \sin \left (\frac{2\pi}{P{_n}}E + \varphi{_n} \right)
    =\frac{a_{n} \sin{i_n}}{c} \left [ \frac{1-e_n^2}{1+e_{n}} \cos f_{n} \sin (f_n+\omega{_n}) \right ] \ ,
	\label{eq:tau}
\end{equation}
where $K_n$ is the amplitude of sinusoidal variation, $P_n$ is the orbital period, $\phi{_n}$ is the orbital phase, $a_{n}\sin{i_n}$ is the projected semi-major axis, $e_n$ is the eccentricity, $f_n$ is the true anomaly of the binary orbit around the centre of the mass of the system, $\omega_n$ is the longitude of the periastron of the $n^{\textup{th}}$ body, and $c$ is the speed of light.

\begin{figure}
	\includegraphics[width=1\columnwidth]{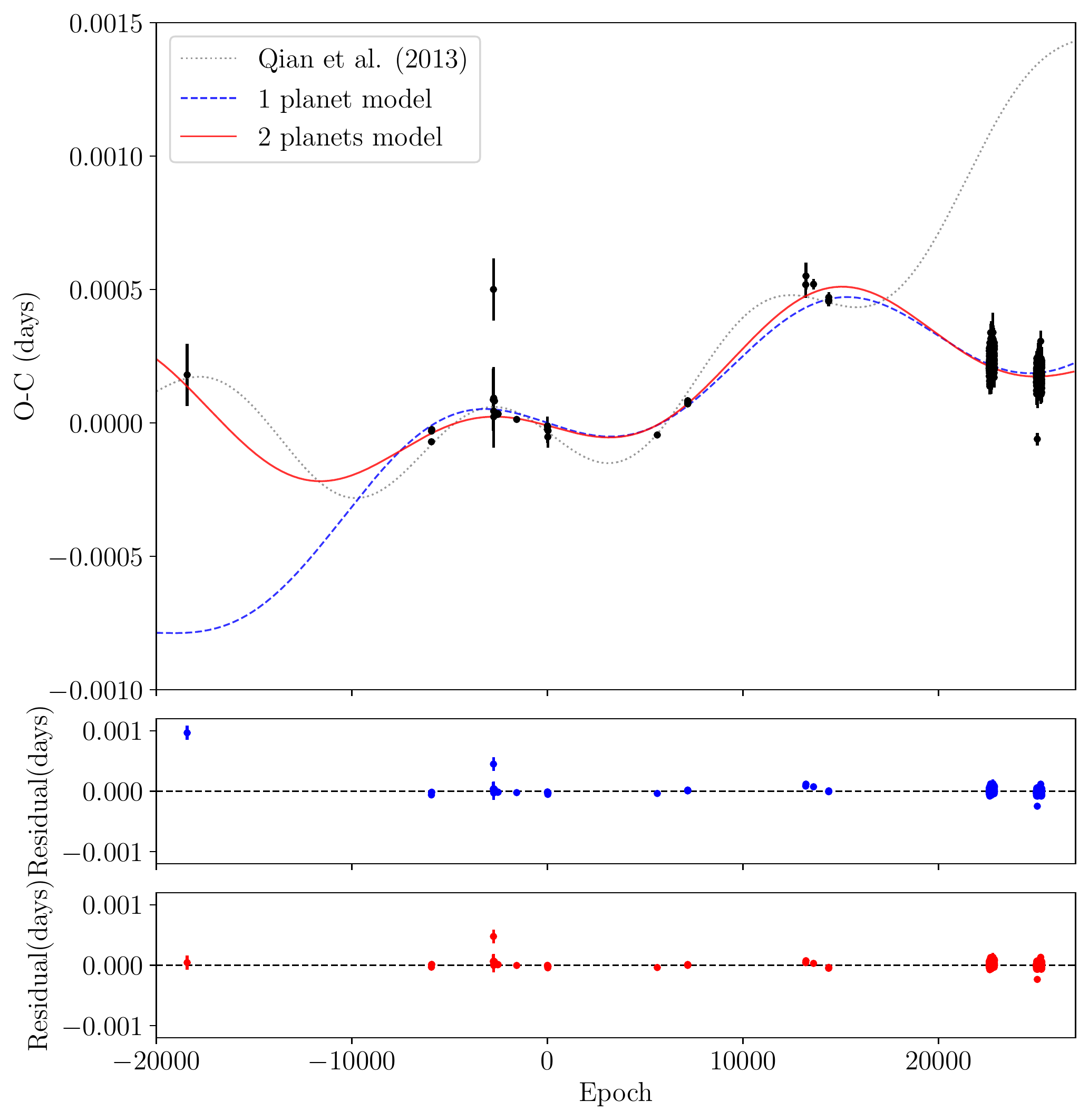}
    \caption{The O-C diagram of RR Cae (Top). The black dotted line shows the \citet{Qian2012} model with a circumbinary planet. The blue dashed line and the red solid line show the MCMC best fitting models of the one circumbinary object model and the two circumbinary objects model, respectively. The residuals of the one circumbinary object model and the two circumbinary objects model are shown in the middle and the bottom panels, respectively.}
    \label{fig:oc}
\end{figure}

A model with one circumbinary object with a circular orbit ($e~=~0$) was fitted with the MCMC sampling of the \texttt{emcee} module \citep{emcee}. The fitting result is shown in Table~\ref{tab:param_2LTT}. The O-C diagram of RR Cae with the modelled time of minima is presented in Figure~\ref{fig:oc}. The one circumbinary object model provides a reduced chi-squared ($\chi^2$) of 8.55. However, the model cannot fit the eclipse time of \cite{Krzeminski1984}. Therefore, the two circumbinary objects model is applied in this work. The model with two circumbinary objects is also shown in Table~\ref{tab:param_2LTT} and Figure~\ref{fig:oc}. The model provides the reduced chi-squared of 4.82 with an outlier at Epoch $E=-2760$. Although the outlier might have affected the LTT fitting, the data are still included in the analyses as the outlier reported by \citet{Bruch1999}. This outlier has already been found in \citet{Parsons2010} work.

\begin{table*}
	\centering
	\caption{The parameters of the circumbinary objects in the RR Cae system.}
	\label{tab:param_2LTT}
	\begin{tabular}{lcccc}
		\hline \hline
		\multirow{2}{*}{Parameters} & \multirow{2}{*}{\citet{Qian2012}} & One object model & \multicolumn{2}{|c|}{Two objects model} \\
		\cline{3-5}
        & & $n = 3$ & $n = 3$ & $n = 4$ \\
        \hline
		$t_0$ (+2 451 523.048 5 d) & $-0.4(\pm0.5)\times 10^{-5}$ & $8.0^{+0.8}_{-0.8}\times 10^{-5}$ & \multicolumn{2}{|c|}{$17^{+7}_{-7}\times 10^{-5}$} \\
		$P$ (+0.303 703 6 d) & $5.79(\pm0.08) \times 10^{-8}$ & $6.5^{+0.1}_{-0.2}\times 10^{-8}$ & \multicolumn{2}{|c|}{$3.9^{+0.5}_{-0.6}\times 10^{-8}$} \\
		$\frac{dP}{dE}$ (d/cycle) & $1.27(\pm0.06) \times 10^{-12}$ & $-1.2^{+0.1}_{-0.1}\times 10^{-12}$ & \multicolumn{2}{|c|}{$0.0^{+0.8}_{-0.7}\times 10^{-12}$}\\
		$\log(K_{n})$ & $-3.78\pm0.13$ & $-3.78^{+0.02}_{-0.02}$ & $-3.86^{+0.04}_{-0.04}$ & $-3.63^{+0.10}_{-0.10}$  \\
		$\omega_{n}$ (rad/epoch) & $4.25(\pm0.02)\times10^{-4}$ & $3.16^{+0.04}_{-0.04}\times 10^{-4}$ & $3.5^{+0.1}_{-0.1}\times 10^{-4}$ & $1.4^{+0.2}_{-0.2}\times 10^{-4}$ \\
		$\varphi_{n}$ (rad) & $2.95\pm0.01$ & $-3.06^{+0.08}_{-0.08}$ & $2.9^{+0.1}_{-0.1}$ & $-0.7^{+0.4}_{-0.4}$ \\
		$P_{n}$ (yr) & $11.9\pm0.1$ & $16.6\pm0.2$ & $15.0\pm0.6$ & $39\pm5$ \\
		$a_{n}$ ($i_n = 90^{\circ}$, AU) & $5.3\pm0.6$ & $5.55\pm0.05$ & $5.2\pm0.1$ & $9.7\pm0.9$ \\
		$M_{n} \sin i_n$ (M$_{\textup{Jup}}$) & $4.2\pm0.4$ & $3.4\pm0.2$ & $3.0\pm0.3$ & $2.7\pm0.7$ \\
		\hline
    \end{tabular}
\end{table*}

From the parameters in Table~\ref{tab:param_2LTT}, the system shows the orbital decay with a rate of $-1.2^{+0.1}_{-0.1}\times 10^{-12}$ d/cycle for the one circumbinary object model and $0.0^{+0.8}_{-0.7}\times 10^{-12}$ d/cycle for the two circumbinary objects model. The cause of the orbital period decreasing in the one circumbinary object model may be due to the angular momentum loss, gravitational radiation, or/and magnetic braking. However, it might be only a part of a long-period cyclic variation, revealing the presence of the second planet in the planetary system \citep{Qian2011}. 

In order to compare the obtained LTT effects with the LLT effect of the planet RR~Cae~b reported by \citet{Qian2012}, the time of minima in Table~\ref{tab:ToM}, which includes published and our newly obtained timing, is used. The model of a circumbinary planet with a planet mass of 4.2~M$_{\textup{Jup}}$ and orbital period of 11.9~yr proposed by \citet{Qian2012} is shown in Figure~\ref{fig:oc}. \citet{Qian2012} LTT model of the planet RR~Cae~b does not fit the time of minima obtained between 2018 and 2020. In comparison, our one circumbinary object model and two circumbinary objects model provide a better fit with the $\chi^2$ of 8.55 and 4.82, respectively.

For the one circumbinary object model, the cyclic variation with a period of $16.6\pm0.2$~yr, which is longer than the period provided by \citet{Qian2012}, is found. The two circumbinary objects model shows that the $a'_{n}\sin{i_n}$ values of the third and fourth components are $0.024\pm0.002$~AU and $0.040\pm0.009$~AU, respectively. The values can infer the orbital periods of $15.0\pm0.6$~yr and $39\pm5$~yr, respectively. The mass function can be written as

\begin{equation}
    f'(m,n)=\frac{(M_{n}\sin{i_{n}})^3}{(M_{1}+M_{2}+\sum_{n=1}^{n} M_{n})^2}=\frac{4\pi^2}{GP_{n}^2} \times (a'_{n}\sin{i_n})^3 \ ,
	\label{eq:mass_fun}
\end{equation}
where $G$ is the gravitational constant. By considering the $M_1~=~0.453~\pm~0.002~\textup{M}_\odot$ and $M_2~=~0.168~\pm~0.001~\textup{M}_\odot$ obtained in Section~\ref{sec:Parameters}, the mass of the third body for the one circumbinary object model is $M_{3}\sin{i_3}$~=~$3.4\pm0.2$~M$_{\textup{Jup}}$. For the two circumbinary objects model, the third and the fourth bodies have the mass function $f(m,3)~=~6\times10^{-8}$~$\textup{M}_\odot$ and $f(m,4)~=~4\times10^{-8}$~$\textup{M}_\odot$. The calculated mass of the third and the fourth bodies are $M_{3}\sin{i_3}$~=~$3.0\pm0.3$~M$_{\textup{Jup}}$ and $M_{4}\sin{i_3}$~=~$2.7\pm0.7$~M$_{\textup{Jup}}$. From the orbital period of $39\pm5$ yr of the fourth body, it might be a known circumbinary planet with the longest orbital period discovered by the eclipse timing technique. Although the two circumbinary objects model provides a better fit for the system’s timing variations, it is hard to conclude that there are two circumbinary objects in the system as the solution of two objects model differs from the solution with one planet by only the eclipse time obtained by \citet{Krzeminski1984}.

\subsection{Planetary orbital stability}

In Section~\ref{sec:LTT}, the one and two circumbinary objects models are proposed. However, the stabilities of the circumbinary objects in both models are concerned. Nowadays, the stability limit for circumbinary planets is not well defined. \citet{Quarles2018} proposed the stability limit based on the calculation of the smallest stable semi-major axis ratio. The numerical tool provided by \citet{Quarles2018} was used to calculate the smallest stable semi-major axis ratio of RR Cae. When $e=0$, the smallest stable ratio between the planetary semi-major axis and the binary semi-major axis is 2.26. Applying the binary parameters obtained in Table~\ref{tab:physical_model}, the critical ratio is 0.017 AU, which is smaller than the semi-major axis of the RR Cae b planet ($a_3 = \sim$5 AU).

\begin{figure}
	\centering
	\includegraphics[width=0.9\columnwidth]{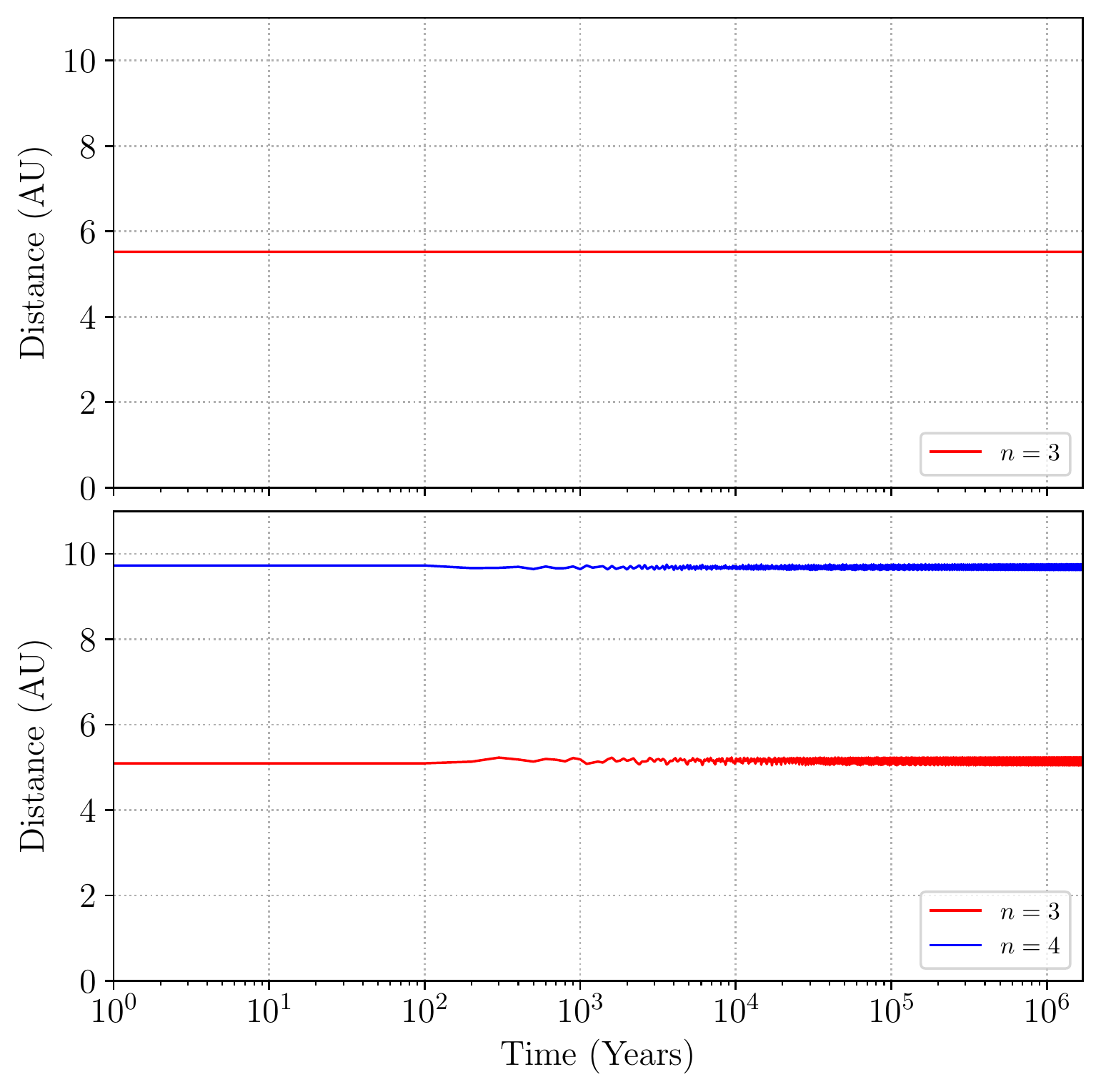}
    \caption{The distances between the circumbinary objects and the system's barycentre as a function of time. The top panel shows the one circumbinary object model and the bottom panel shows the two circumbinary objects model. The models are simulated using the \texttt{REBOUND} package.}
    \label{fig:stable}
\end{figure}

Nevertheless, as the two circumbinary objects model is proposed in this work, the stability of the two circumbinary objects model is much more complex than the model with one circumbinary object. In order to confirm their stability, the systems are simulated using the \texttt{REBOUND} package, an N-body integrator \citep{REBOUND}. In this work, we are only interested in short-term orbital stability. Due to the limit of computational time, in both one and two circumbinary objects models, a 1.7 million years orbit ($\sim$1$\%$ of the system age obtained in Section~\ref{sec:Parameters}) with a step size of 1 year is simulated with the parameters of the binary and circumbinary object listed in Tables~\ref{tab:physical_model} and~\ref{tab:param_2LTT}. The planetary Roche limit and mutual Hill's sphere are assigned to be orbital stability limits. The distances between the circumbinary objects and the system barycentre as a function of time are shown in Figure~\ref{fig:stable}.

In the one circumbinary object model, the binary does not show any strong gravitational interaction on the circumbinary object. For the two circumbinary objects model, the perturbation can be found in both objects in the timescale of hundreds of years. This timescale corresponds to the period of two proposed circumbinary objects. In the timeframe of a million years, both objects have no significant sign of instability. Therefore, the stability of both models can be confirmed.

\subsection{The Applegate mechanism}

In Section~\ref{sec:LTT}, the eclipsing timing variation of RR Cae is well-fitted by one or two circumbinary objects models. However, the LTT effect from circumbinary objects is not the only known cause of the cyclic timing variation of an eclipsing binary in the scale between 10-100~yr. Other mechanisms, such as the Applegate mechanism \citep{Applegate1992}, can also explain this cyclic variation. The Applegate mechanism results from quasiperiodic changes in the quadrupole moment of a magnetically active component, the secondary component in this work, due to the radial distribution of angular momentum inside the component. The magnetic activity cycle can lead to cyclic variations of the binary orbital period. Therefore, in order to confirm whether the variation is driven by the Applegate mechanism or not, the effects of the Applegate mechanism are tested.

The obtained cyclic variation parameters of RR Cae (Table~\ref{tab:param_2LTT}) are used to calculate the magnetic energy provided by its secondary component. This energy was mathematically estimated by \citet{Applegate1992} and refined by \citet{Brinkworth2006}. Recently, \citet{Volschow2016} extended the \citet{Brinkworth2006} estimation by including quadrupole moment changes in two zones: the core and the shell. Using the Applegate calculator based on the two-zone model of \citet{Volschow2016}\footnote{Applegate calculator: \texttt{http://www.theory-starformation-group.cl /applegate/}}, the required energy to drive the Applegate mechanism of one circumbinary object model is 20 times the available energy in the active M-dwarf. The required energies for the two circumbinary objects model are 18 and 3.5 times higher for $n=3$ and $n=4$, respectively.

\subsection{Effects of out-of-eclipse slope}

Eclipse timing variations in an eclipsing binary might be caused by not only the circumbinary objects or the Applegate mechanism but also the starspot on a binary component \citep{Watson2004,Conroy2014,Mazeh2015}. \citet{Mazeh2015} showed that the eclipse timing variation could be induced by the out-of-eclipse light curve slope caused by the presence of a starspot and provided the analytic approximation of the variation. We applied their analytical approximation to the RR Cae model with starspot. We found that the timing variation of the system corresponding to the out-of-eclipse slope is less than $10^{-5}$ d. Comparing this timescale with the amplitude of the variation in the O-C diagram ($\sim10^{-4}$ d), it can be confirmed that the out-of-eclipse slope is not the primary source of the obtained O-C variation.

The presence of a starspot should not cause the timing variation on the light curve fitted with \texttt{PHOEBE}, as the out-of-eclipse variation corresponding to the spot is already included in the fitting model. In order to satisfy the assumption, the RR Cae light curves with a large bright spot obtained from the TESS 2018 data, as in Table~\ref{tab:spot_model}, are simulated with six different spot longitudes ($\lambda$): 0, 60, 120, 180, 240, and 300 degrees. The 251 data points ($\sim$52 s cadence) light curves with phases between -0.25 and 0.25 are created, and MCMC is fitted with two \texttt{PHOEBE} models: models with and without a starspot. 

\begin{figure}
	\includegraphics[width=1.02\columnwidth]{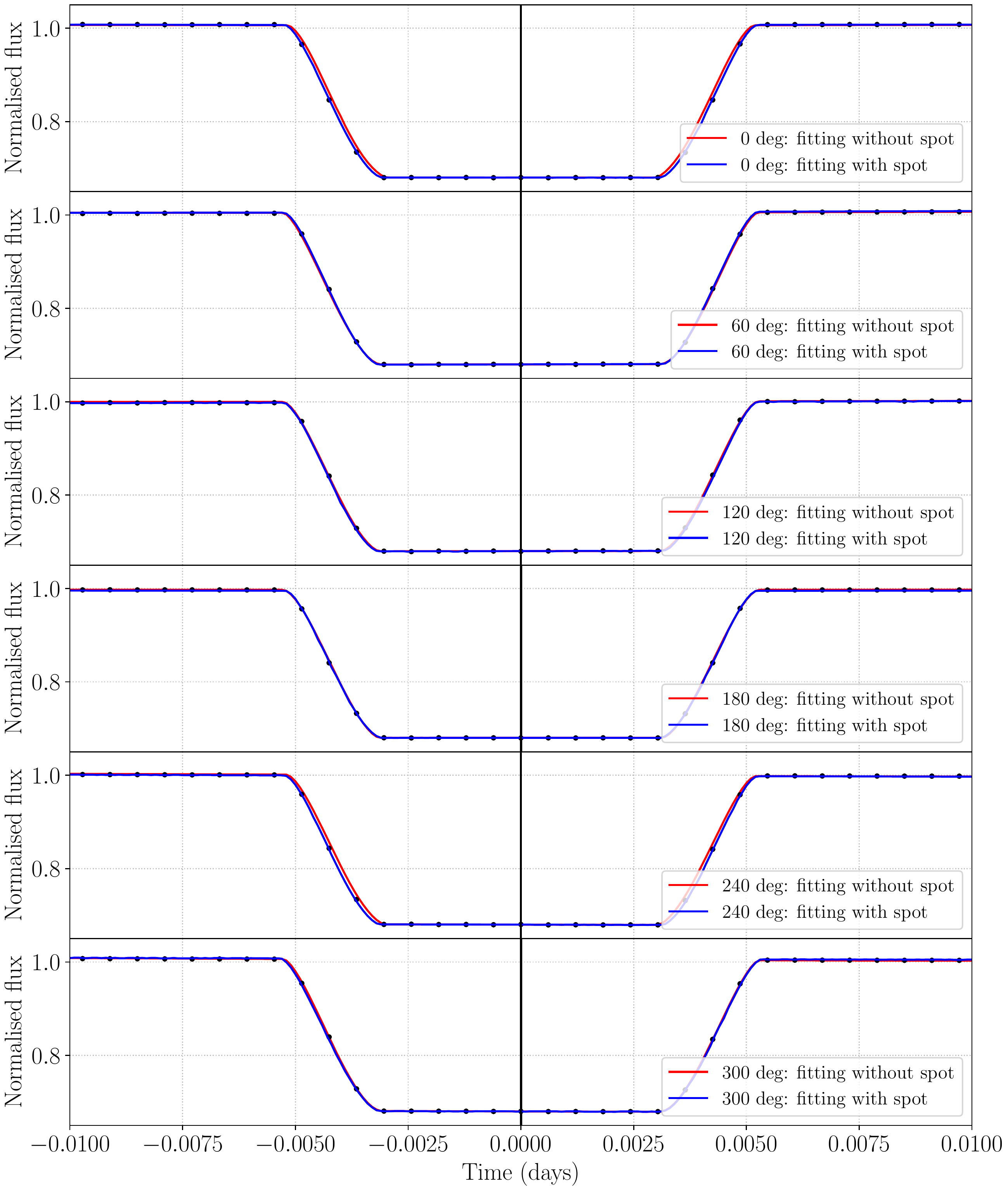}
    \caption{The simulated TESS 2018 light curves with six different starspot longitudes fitted with the \texttt{PHOEBE} models without a starspot (Red) and with a starspot (Blue).}
    \label{fig:spotvar}
\end{figure}

\begin{figure}
	\centering
	\includegraphics[width=0.9\columnwidth]{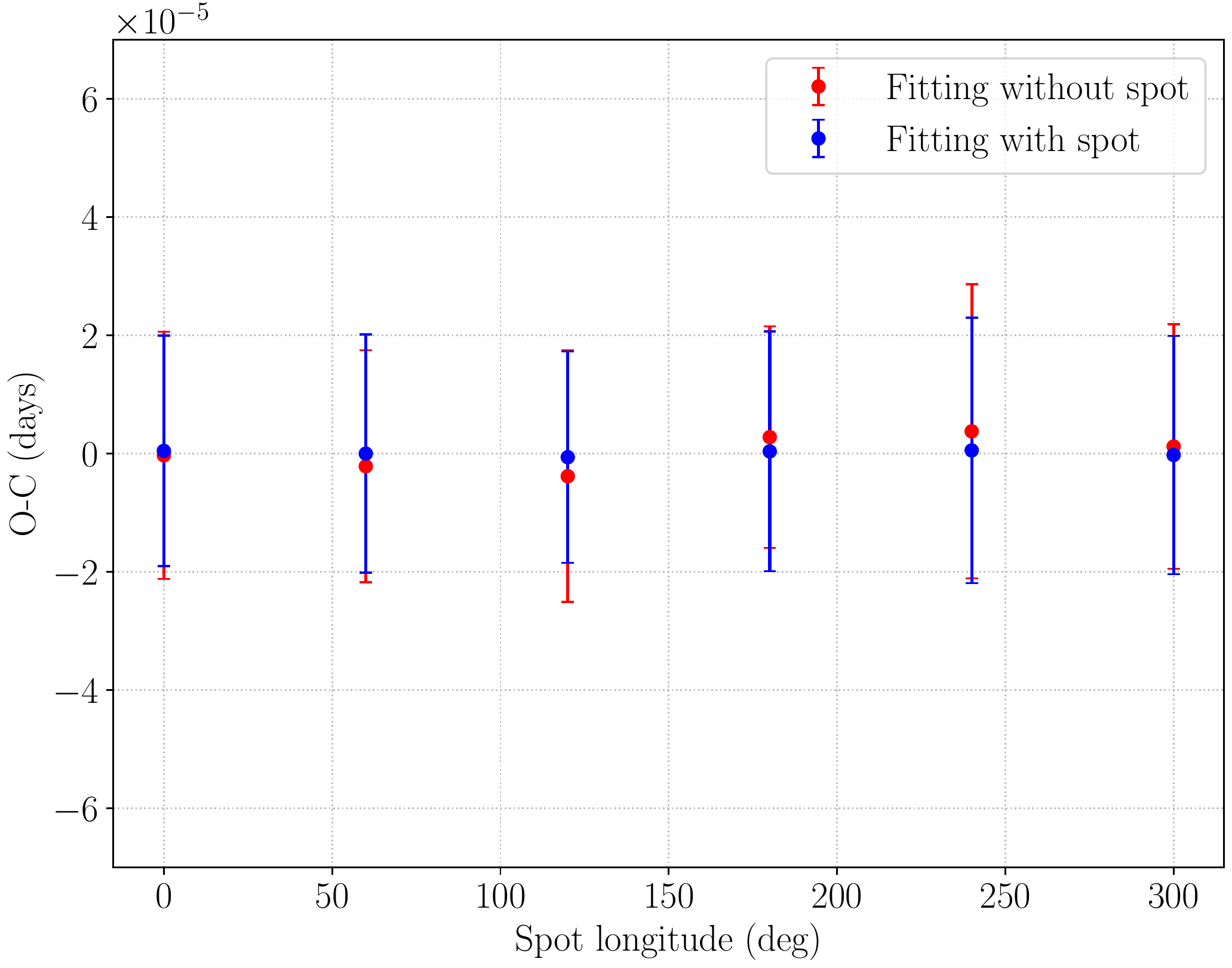}
    \caption{The eclipse timing variations obtained from the best fitting models without spot (Red) and with spot (Blue) on the simulated TESS 2018 light curves with six different spot longitudes using the \texttt{PHOEBE} package.}
    \label{fig:spotresult}
\end{figure}

Figure~\ref{fig:spotvar} shows simulated light curves with the two fitting models. As discussed in Section~\ref{sec:Parameters}, the model with starspot provides a better fitting. In Figure~\ref{fig:spotresult}, the eclipse timing variations of both fitting models in various spot longitudes are compared. The fitting from the model without a spot provides a higher timing variation due to the effect of the out-of-eclipse slope, as proposed by \citet{Mazeh2015}. However, their timing variations are still very small ($\sim10^{-6}$~d) compared to the system O-C variations ($\sim10^{-4}$~d). In conclusion, if there is no very hot or cold spot that caused a strong out-of-eclipse variation in the past decades, the out-of-eclipse slope is not the main driving source of the obtained O-C variations.

\section{Conclusions}
\label{sec:Conclusion}

In this work, we observed and studied a detached binary system, RR~Cae. The system has a circumbinary planet, RR~Cae~b, which is a 4.2 Jupiter mass planet with an orbital period of 11.9~yr. The planet was discovered by \citet{Qian2012} using the light travel time effect. Between 2018 and 2020, optical multi-filter observations of the binary were obtained with the TESS, the 0.6-m PROMPT-8 telescope in R-band, and the 0.7-m Thai Robotic Telescope at Spring Brook Observatory in the R- and the I- bands. A total of 430 primary eclipses were obtained.

In order to revise the RR Cae parameters, we use our photometric data combined with the spectra from the VLT (Program ID: 076.D-0142, PI: Maxted, Pierre). From the VLT data, there are only three spectral lines: H$_\alpha$, Ca II (K line), and Ca II (H line), which can be used to calculate radial velocity solutions for both primary and secondary components. These three spectral lines provide an average gravitational redshift of around $16.2~\pm~0.4$~km~s$^{-1}$. However, only the H$_\alpha$ spectral lines are used in this work as the highest signal-to-noise spectral lines. 

The data are analysed with the \texttt{PHOEBE} code \citep{Prsa2016} and performed the fitting with the MCMC method. The TESS light curves required the model with a starspot in order to fit out-of-eclipse variation. The model with only a starspot is used in this work. The derived data from the analysis provide an orbital period $P=0.303 703 61^{+1\textup{E}-8}_{-2\textup{E}-8}$~d, orbital inclination $i=82.9^{+0.2}_{-0.3}$~deg and mass ratio $q=0.371^{+0.002}_{-0.002}$, which correspond to the primary component mass, $M_1$, of $0.453^{+0.002}_{-0.002}~\textup{M}_\odot$ and the secondary component mass, $M_2$, of $0.168^{+0.001}_{-0.001}~\textup{M}_\odot$. The TESS data show that the light curves in 2018 and 2020 have different out-of-eclipse variations. Therefore, different spot models on the secondary component are adopted to fit the TESS data in 2018 and 2020. The TESS data in 2018 show a hot spot near the stellar pole. However, a cold spot at a moderate latitude is preferred to the data in 2020. The presence of this large spot might be explained by the heat transfer between the two components.

The eclipse timing variation of RR Cae shows cyclic variations in the observed timing. We fitted the eclipse timing variation with a binary's period change model with the light travel time effect from one or two circumbinary objects. The one circumbinary object model shows a cyclic variation of $16.6\pm0.2$~yr with an amplitude of $14\pm1$~s, corresponding to a planet with a minimum mass of $3.4~\pm~0.2$~M$_{\textup{Jup}}$. The one circumbinary object fitting model provides the reduced chi-squared value of 8.55. In comparison, the two circumbinary objects provide a better fit with the reduced chi-squared value of 4.82. The two circumbinary objects fitting model shows that RR Cae has periodic variations with amplitudes of $12\pm1$~s and $20\pm5$~s with periods of $15.0\pm0.5$~yr and $39\pm5$~yr, respectively. These variations can be caused by two circumbinary planets with masses of $M_{3}\sin{i_3}$~=~$3.0\pm0.3$~M$_{\textup{Jup}}$ and $M_{4}\sin{i_4}$~=~$2.7\pm0.7$~M$_{\textup{Jup}}$.

In order to prove that the variation due to the third and the fourth bodies in the system is sensible, the stability of the bodies is confirmed with analytic calculation by \citet{Quarles2018} and simulation with the \texttt{REBOUND} package \citep{REBOUND}. Although the LTT effect due to the gravitational interaction of circumbinary objects is proposed, many other effects can still cause the eclipsing timing variation. The effects of the Applegate mechanism and the effects of out-of-eclipse slope on the timing variations are tested to confirm that these two effects are not the primary driving source of the timing variation of the system. However, to ensure the second planet’s presence in the system, long-term continuous monitoring of RR Cae is needed as the solution of two objects differs from the solution with one planet by only the eclipse time obtained by \citet{Krzeminski1984}.

\section*{Acknowledgements}

The authors acknowledge the anonymous referees for their valuable suggestions that helped to improve the paper. This work is based on observations made with the Thai Robotic Telescope, which is operated by the National Astronomical Research Institute of Thailand (Public Organization), and the TESS data of RR Cae from MAST: Barbara A. Mikulski Archive for Space Telescopes. This work is based on data products from observations made with ESO Telescopes at the La Silla Paranal Observatory under program ID 076.D-0142.

This work presents results from the European Space Agency (ESA) space mission Gaia. Gaia data are being processed by the Gaia Data Processing and Analysis Consortium (DPAC). Funding for the DPAC is provided by national institutions, in particular the institutions participating in the Gaia MultiLateral Agreement (MLA). The Gaia mission website is \texttt{https://www.cosmos.esa.int/gaia}. The Gaia archive website is \texttt{https://archives.esac.esa.int/gaia}.

This work is also supported by a National Astronomical Research Institute of Thailand (NARIT) research grant. This work has been done using the facilities of Chalawan High-Performance Computing at the National Astronomical Research Institute of Thailand (NARIT).

\section*{Data availability}
The PROMPT-8, and the TRT-SBO photometric data, the VLT radial velocities data, and the primary eclipse time of minima of RR Cae are available in the article and in the online supplementary material.


\bibliographystyle{mnras}
\bibliography{RRCae}

\bsp	
\label{lastpage}

\appendix

\section{RR Cae light curves from the PROMPT-8 telescope and TRT-SBO.}
\label{sec:App_LC}

\begin{figure}
	\includegraphics[width=0.9\columnwidth]{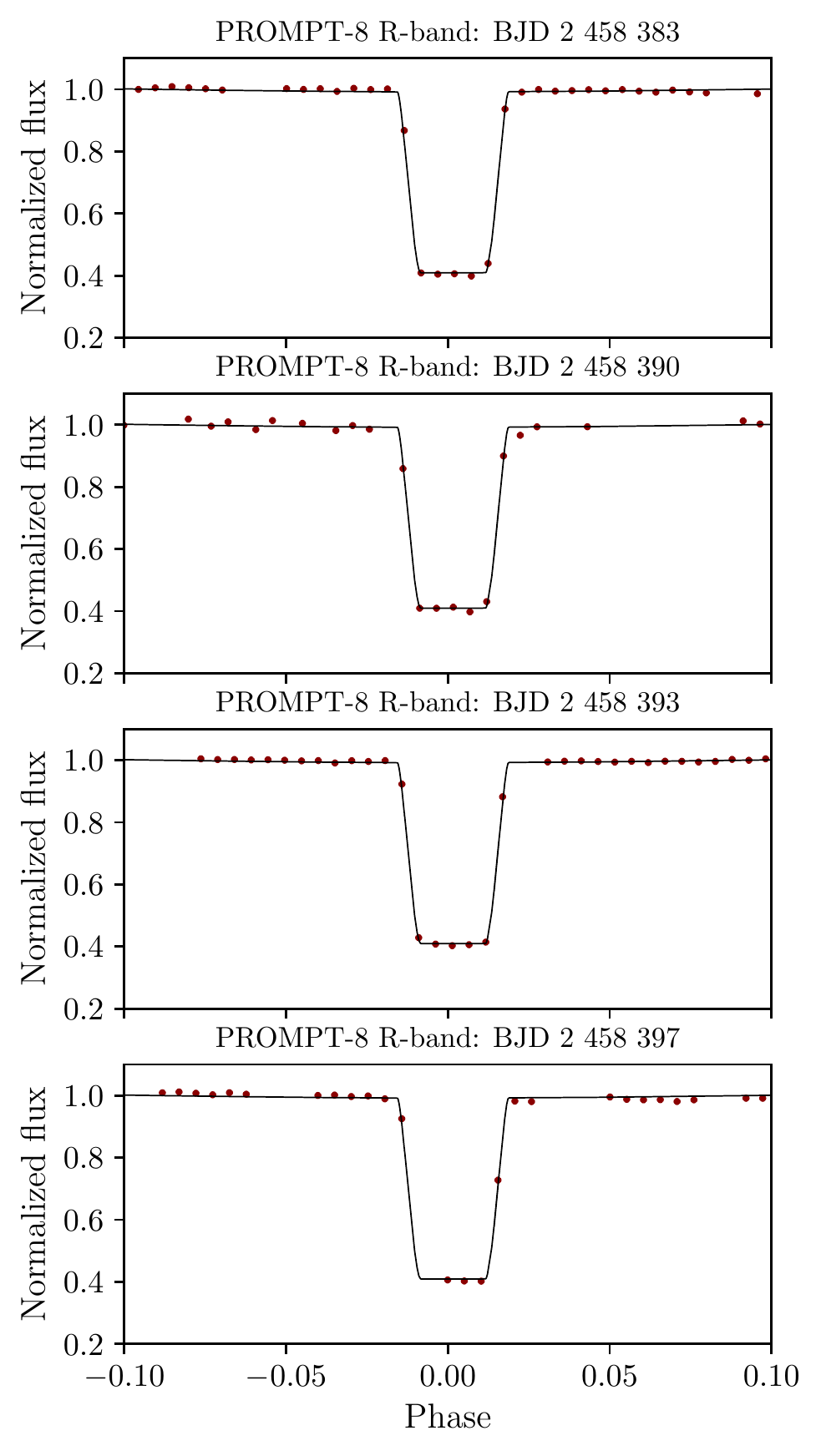}
    \caption{RR Cae light curves obtained from the PROMPT-8 telescope in R-band with the best fitting model from the \texttt{PHOEBE} code.}
\end{figure}

\begin{figure}
	\includegraphics[width=0.9\columnwidth]{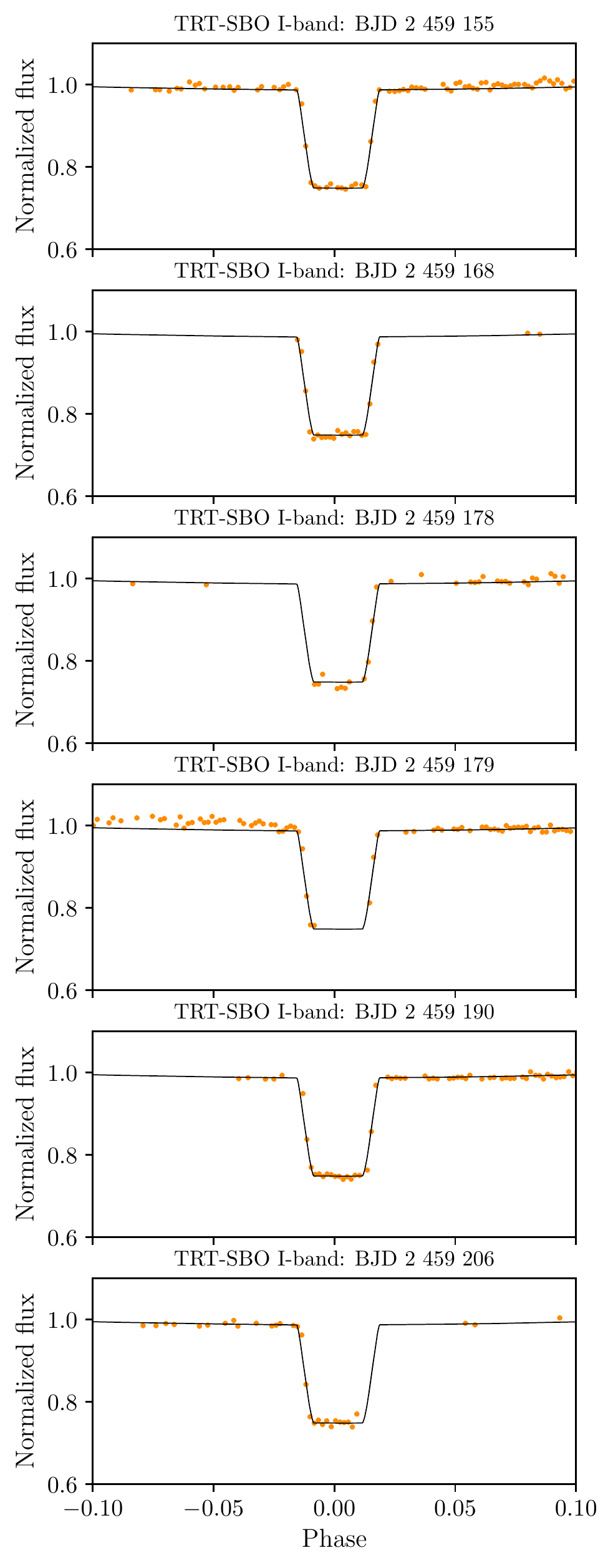}
    \caption{RR Cae light curves obtained from the TRT-SBO telescope in I-band with the best fitting model from the \texttt{PHOEBE} code.}
\end{figure}

\begin{figure*}
	\includegraphics[width=1.8\columnwidth]{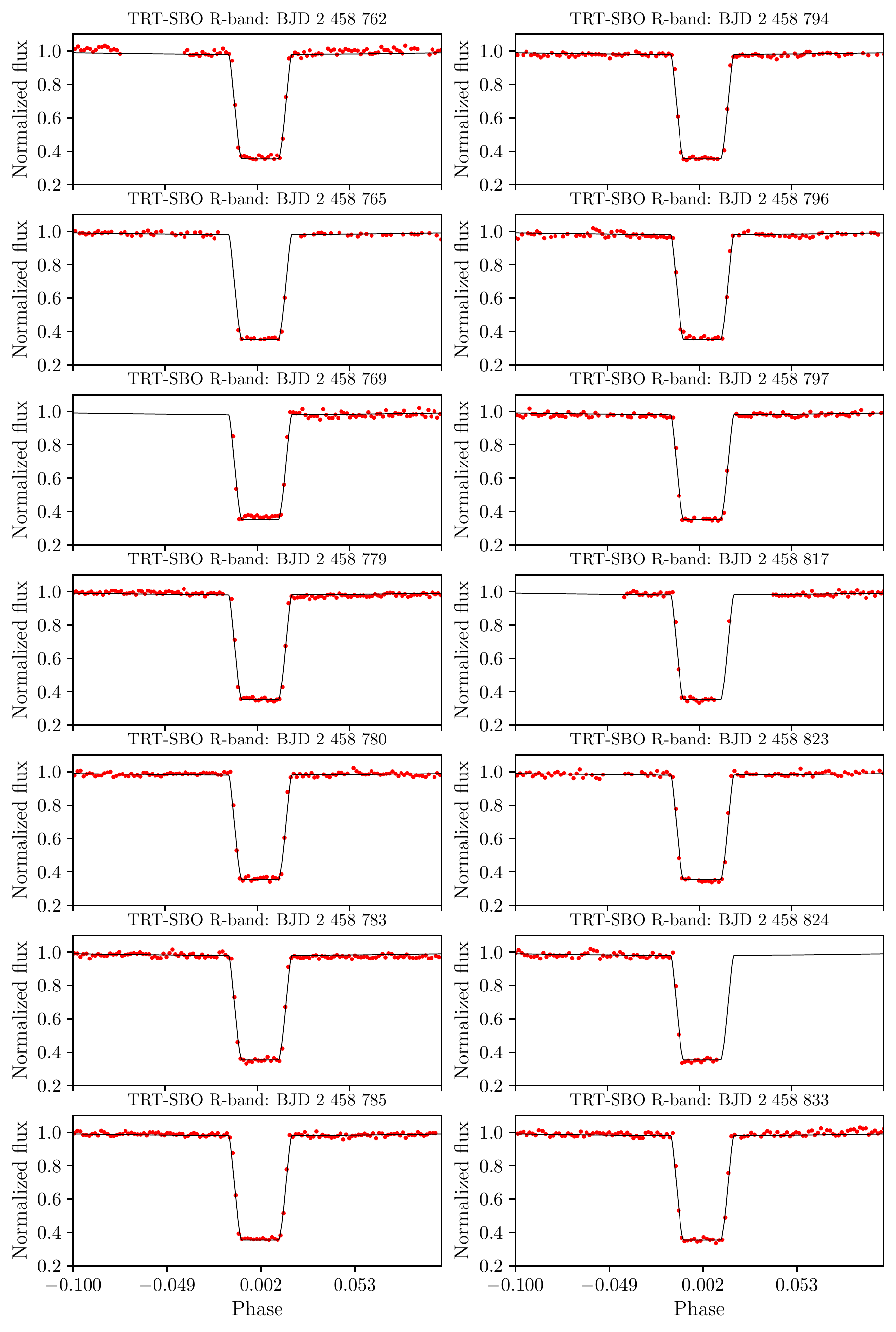}
    \caption{RR Cae light curves obtained from the TRT-SBO telescope in R-band with the best fitting model from the \texttt{PHOEBE} code.}
\end{figure*}

\section{Radial velocities of the RR Cae spectra from the UVES/VLT.}

\onecolumn
\begin{landscape}
    \begin{table*}
	    \caption{Radial velocities with errors of the RR Cae spectra from the UVES/VLT.}
	    \label{tab:RV_results}
	    \begin{center}
        \begin{tabular}{ccrrrrrrrrrr}
	        \hline \hline
            \multirow{2}{*}{BJD} & \multirow{2}{*}{Phase} &         \multicolumn{10}{c}{Radial velocity and error values (km s$^{-1}$)} \\
            \cline{3-12}
            &  & $\textup{H}_{\alpha,1,\textup{emi}}$  & $\textup{H}_{\alpha,2,\textup{emi}}$ & $\textup{H}_{\beta,2,\textup{emi}}$ & $\textup{H}_{\gamma,2,\textup{emi}}$ & $\textup{H}_{\delta,2,\textup{emi}}$ & $\textup{CaII(H)}_{1,\textup{abs}}$ & $\textup{CaII(H)}_{2,\textup{emi}}$ & $\textup{CaII(K)}_{1,\textup{abs}}$ & $\textup{CaII(K)}_{2,\textup{emi}}$ & $\textup{CaI}_{1,\textup{abs}}$ \\
	        \hline
	        2 453 658.756 04 & 0.209 10 & 15.7 $\pm$ 0.9 & 261.8 $\pm$ 0.2 & 293.7 $\pm$ 0.3 & 306.8 $\pm$ 0.5 & 256.3 $\pm$ 0.7 & 61.1 $\pm$ 0.3 & 310.2 $\pm$ 0.8 & 1.0 $\pm$ 0.5 & 248.1 $\pm$ 0.4 & 7.7 $\pm$ 0.4 \\
	        2 453 658.760 87 & 0.224 99 & 13.9 $\pm$ 0.9 & 266.2 $\pm$ 0.2 & 298.7 $\pm$ 0.3 & 312.7 $\pm$ 0.5 & 258.9 $\pm$ 0.7 & 59.5 $\pm$ 0.3 & 314.4 $\pm$ 0.8 & -1.6 $\pm$ 0.4 & 250.8 $\pm$ 0.4 & 6.2 $\pm$ 0.5 \\
	        2 453 658.765 61 & 0.240 59 & 13.4 $\pm$ 0.8 & 268.1 $\pm$ 0.1 & 299.8 $\pm$ 0.3 & 313.2 $\pm$ 0.5 & 259.2 $\pm$ 0.8 & 58.7 $\pm$ 0.2 & 314.9 $\pm$ 0.8 & -1.6 $\pm$ 0.4 & 252.3 $\pm$ 0.4 & 6.0 $\pm$ 0.5 \\
	        2 453 658.770 43 & 0.256 46 & 13.9 $\pm$ 0.8 & 267.9 $\pm$ 0.1 & 298.6 $\pm$ 0.3 & 312.7 $\pm$ 0.6 & 259.8 $\pm$ 0.9 & 59.4 $\pm$ 0.3 & 315.2 $\pm$ 0.8 & -1.2 $\pm$ 0.5 & 252.1 $\pm$ 0.4 & 4.9 $\pm$ 0.4 \\
	        2 453 658.775 23 & 0.272 28 & 14.4 $\pm$ 0.9 & 266.4 $\pm$ 0.2 & 297.2 $\pm$ 0.3 & 312.2 $\pm$ 0.6 & 256.8 $\pm$ 0.8 & 59.9 $\pm$ 0.4 & 314.2 $\pm$ 1.0 &  -0.7 $\pm$ 0.4 & 249.5 $\pm$ 0.5 & 5.6 $\pm$ 0.4\\
	        2 453 658.780 05 & 0.288 15 & 15.8 $\pm$ 0.8 & 262.7 $\pm$ 0.2 & 293.8 $\pm$ 0.3 & 308.5 $\pm$ 0.6 & 253.0 $\pm$ 1.0 & 61.6 $\pm$ 0.3 & 310.5 $\pm$ 1.0 &  0.7 $\pm$ 0.5 & 247.5 $\pm$ 0.5 & 8.3 $\pm$ 0.4 \\
	        2 453 658.784 88 & 0.304 06 & 18.8 $\pm$ 0.8 &  256.8 $\pm$ 0.2 & 287.4 $\pm$ 0.3 & 301.8 $\pm$ 0.7 & 246.1 $\pm$ 1.0 & 63.7 $\pm$ 0.3 & 301.8 $\pm$ 1.3 & 2.8  $\pm$ 0.5 & 240.6 $\pm$ 0.5 & 10.7 $\pm$ 0.4 \\
	        2 453 658.789 70 & 0.319 92 & 21.3 $\pm$ 0.6 & 249.5 $\pm$ 0.2 & 279.7 $\pm$ 0.3 & 293.9 $\pm$ 0.7 & 239.0 $\pm$ 1.1 & 67.1 $\pm$ 0.4 & 298.2 $\pm$ 1.4 &  5.7 $\pm$ 0.4 & 233.7 $\pm$ 0.5 & 13.4 $\pm$ 0.4 \\
	        2 453 659.666 52 & 0.207 01 &  15.1 $\pm$ 1.0 & 255.0 $\pm$ 0.1 & 284.3 $\pm$ 0.4  &  298.0 $\pm$ 0.6 & 246.0 $\pm$ 0.8 & 62.6 $\pm$ 0.4 & 298.5 $\pm$ 1.0 & 1.6 $\pm$ 0.5 & 238.6 $\pm$ 0.4 &  8.1 $\pm$ 0.4 \\
	    
	        2 453 659.671 14 & 0.222 21 &  14.0 $\pm$ 1.1 & 260.6 $\pm$ 0.1 & 291.5 $\pm$ 0.3 &  300.7 $\pm$ 0.6 & 253.7 $\pm$ 0.8 & 60.8 $\pm$ 0.3 & 307.3 $\pm$ 1.1 & -1.1 $\pm$ 0.5 & 246.0 $\pm$ 0.4 &  7.5 $\pm$ 0.5 \\
	    
	        2 453 659.675 96 & 0.238 11 &  10.7 $\pm$ 0.9 & 263.2 $\pm$ 0.1 & 293.6 $\pm$ 0.3 &  307.9 $\pm$ 0.6 & 254.1 $\pm$ 0.8 & 60.1 $\pm$ 0.2 & 310.5 $\pm$ 1.0 & -1.6 $\pm$ 0.5 & 249.4 $\pm$ 0.5 & 6.4 $\pm$ 0.5 \\
	    
	        2 453 659.680 86 & 0.254 22 &  12.9 $\pm$ 0.9 & 263.8 $\pm$ 0.1 & 293.2 $\pm$ 0.3 &  308.5 $\pm$ 0.7 & 254.4 $\pm$ 0.8 & 59.8 $\pm$ 0.3 & 309.9 $\pm$ 0.9 & -1.1 $\pm$ 0.5 & 249.2 $\pm$ 0.4 & 5.5 $\pm$ 0.5 \\
	    
	        2 453 659.685 68 & 0.270 10 &  12.6 $\pm$ 1.0 & 262.8 $\pm$ 0.2 & 292.5 $\pm$ 0.3 &  308.0 $\pm$ 0.7 & 251.5 $\pm$ 1.1 & 60.4 $\pm$ 0.3 & 309.8 $\pm$ 0.8 & -1.0 $\pm$ 0.4 & 247.3 $\pm$ 0.4 & 6.7 $\pm$ 0.5 \\
	    
	        2 453 659.690 69 & 0.286 58 &  13.9 $\pm$ 1.1 & 259.8 $\pm$ 0.2 & 290.3 $\pm$ 0.4 &  305.2 $\pm$ 0.6 & 250.5 $\pm$ 0.9 & 61.3 $\pm$ 0.2 & 307.2 $\pm$ 0.8 & -0.2 $\pm$ 0.4 & 245.2 $\pm$ 0.4 & 7.9 $\pm$ 0.4 \\
	    
	        2 453 659.695 41 & 0.302 13 &  16.6 $\pm$ 1.0 & 255.2 $\pm$ 0.1 & 286.7 $\pm$ 0.4 &  301.9 $\pm$ 0.5 & 248.9 $\pm$ 0.9 & 62.1 $\pm$ 0.3 & 304.1 $\pm$ 0.6 & 2.5 $\pm$ 0.5 & 240.3 $\pm$ 0.3 & 9.7 $\pm$ 0.4 \\
	    
	        2 453 659.700 30 & 0.318 23 &  18.7 $\pm$ 0.9 & 248.2 $\pm$ 0.2 & 279.6 $\pm$ 0.3 &  295.2 $\pm$ 0.5 & 240.7 $\pm$ 0.9 & 66.7 $\pm$ 0.3 & 295.8 $\pm$ 0.7 & 5.3 $\pm$ 0.5 & 234.5 $\pm$ 0.4 & 11.4 $\pm$ 0.4 \\
	    
	        2 453 660.762 88 & 0.816 97 &  152.7 $\pm$ 0.6 & -106.3 $\pm$ 0.2 & -75.7 $\pm$ 0.4 &  -62.8 $\pm$ 0.6 & -112.9 $\pm$ 0.9 & 199.6 $\pm$ 0.3 & -58.1 $\pm$ 0.6 & 138.5 $\pm$ 0.5 & -121.8 $\pm$ 0.5 & 145.4 $\pm$ 0.4 \\
	    
	        2 453 660.768 04 & 0.833 96 & 150.1 $\pm$ 0.6 & -97.9 $\pm$ 0.2 & -66.9 $\pm$ 0.3 &  -53.0 $\pm$ 0.6 & -105.6 $\pm$ 1.4 & 195.8 $\pm$ 0.3 & -51.1 $\pm$ 0.6 & 134.6 $\pm$ 0.6 & -112.1 $\pm$ 0.4 & 141.2 $\pm$ 0.5 \\
	    
	        2 453 660.773 05 & 0.850 45 &  147.1 $\pm$ 0.5 & -86.8 $\pm$ 0.2 & -56.4 $\pm$ 0.3 &  -42.7 $\pm$ 0.5 & -95.3 $\pm$ 1.3 & 192.4 $\pm$ 0.4 & -41.0 $\pm$ 0.5 & 131.0 $\pm$ 0.4 & -99.4 $\pm$ 0.3 & 138.8 $\pm$ 0.5 \\
	    
	        2 453 660.778 08 & 0.867 01 &  142.5 $\pm$ 0.7 & -73.5 $\pm$ 0.2 & -43.6 $\pm$ 0.3 &  -30.3 $\pm$ 0.5 & -82.0 $\pm$ 1.4 & 187.8 $\pm$ 0.4 & -27.0 $\pm$ 0.8 & 126.0 $\pm$ 0.4 & -88.2 $\pm$ 0.4 & 133.0 $\pm$ 0.5 \\
	    
	        2 453 660.783 09 & 0.883 53 & 136.3 $\pm$ 0.7 & -59.1 $\pm$ 0.2 & -29.6 $\pm$ 0.5 & -16.4 $\pm$ 0.6 & -66.5 $\pm$ 1.3 & 182.9 $\pm$ 0.4 & -12.4 $\pm$ 0.7 & 120.9 $\pm$ 0.5 & -74.5 $\pm$ 0.5 & 128.8 $\pm$ 0.5 \\
	    
	        2 453 660.788 12 & 0.900 07 & 130.5 $\pm$ 0.6 & -43.1 $\pm$ 0.2 & -13.4 $\pm$ 1.0 &  0.5 $\pm$ 0.7 & -51.5 $\pm$ 1.4 & 177.2 $\pm$ 0.4 & 5.1 $\pm$ 0.7 & 116.3 $\pm$ 0.4 & -57.4 $\pm$ 0.7 & 122.9 $\pm$ 0.6 \\
	    
	        2 453 660.793 14 & 0.916 63 &  123.7 $\pm$ 0.6 & -26.2 $\pm$ 0.4 & 4.4 $\pm$ 0.4 &  17.6 $\pm$ 0.7 & -35.1 $\pm$ 1.2 &  172.1 $\pm$ 0.5 & 20.2 $\pm$ 0.9 & 109.3 $\pm$ 0.4 & -41.3 $\pm$ 0.6 & 116.5 $\pm$ 0.5 \\
	    
	        2 453 660.798 18 & 0.933 20 & 117.1 $\pm$ 0.7 & -8.5 $\pm$ 0.3 & 21.0 $\pm$ 0.5 &  34.7 $\pm$ 0.8 & -20.2 $\pm$ 0.8 & 164.6 $\pm$ 0.5 & 35.2 $\pm$ 0.8 & 102.3 $\pm$ 0.7 & -24.1 $\pm$ 0.5 & 109.0 $\pm$ 0.5 \\
	    
	        2 453 661.736 68 & 0.023 39 & -- &  99.2 $\pm$ 0.2 & 131.0 $\pm$ 0.5 &  144.2 $\pm$ 0.6 & 94.3 $\pm$ 1.3 & 107.6 $\pm$ 0.6 & 151.7 $\pm$ 0.6 & 40.6 $\pm$ 0.5 & 93.4 $\pm$ 0.5 & 63.0 $\pm$ 0.5 \\
	    
	        2 453 661.741 77 & 0.040 14 & -- &  120.3 $\pm$ 0.2 & 151.9 $\pm$ 0.4 &  162.6 $\pm$ 0.7 & 113.9 $\pm$ 1.1 & 102.5 $\pm$ 0.3 & 166.4 $\pm$ 1.1 & 34.4 $\pm$ 0.5 & 107.1 $\pm$ 0.4 & 56.7 $\pm$ 0.5 \\

	        2 453 661.746 79 & 0.056 68 & -- &  141.1 $\pm$ 0.2 & 172.9 $\pm$ 0.4 & 178.3 $\pm$ 0.6 & 130.6 $\pm$ 1.1 & 96.2 $\pm$ 0.3 & 186.2 $\pm$ 0.9 & 28.8 $\pm$ 0.5 & 124.4 $\pm$ 0.5 & 49.7 $\pm$ 0.5 \\
	    
	        2 453 661.751 80 & 0.073 19 &  56.2 $\pm$ 0.7 & 158.3 $\pm$ 0.2 & 189.4 $\pm$ 0.5 &  196.0 $\pm$ 0.6 & 146.0 $\pm$ 1.2 & 90.7 $\pm$ 0.3 & 201.3 $\pm$ 0.9 & 21.9 $\pm$ 0.4 & 140.3 $\pm$ 0.5 & 42.4 $\pm$ 0.5 \\
	    
	        2 453 661.756 86 & 0.089 83 &  44.4 $\pm$ 0.7 & 174.8 $\pm$ 0.2 & 206.2 $\pm$ 0.4 &  214.7 $\pm$ 0.7 & 164.1 $\pm$ 1.1 & 82.9 $\pm$ 0.3 & 221.1 $\pm$ 1.2 & 16.1 $\pm$ 0.6 & 156.1 $\pm$ 0.5 & 35.8 $\pm$ 0.4 \\
	    
	        2 453 661.761 87 & 0.106 34 &  40.3 $\pm$ 0.6 & 190.6 $\pm$ 0.2 & 222.2 $\pm$ 0.4 &  229.2 $\pm$ 0.7 & 180.5 $\pm$ 1.4 & 77.4 $\pm$ 0.3 & 236.8 $\pm$ 1.2  & -- & 173.6 $\pm$ 0.7 & 30.1 $\pm$ 0.4 \\
	    
	        2 453 661.766 90 & 0.122 88 & 37.7 $\pm$ 0.6 & 205.5 $\pm$ 0.2 & 237.5 $\pm$ 0.4 &  243.8 $\pm$ 0.7 & 197.0 $\pm$ 1.0 & 72.7 $\pm$ 0.3 & 251.2 $\pm$ 0.9 & -- & 188.5 $\pm$ 0.5 & 24.5 $\pm$ 0.5 \\
	    
	        2 453 662.771 91 & 0.139 39 & 33.1 $\pm$ 0.7 & 219.8 $\pm$ 0.2 & 251.7 $\pm$ 0.3 & 259.6 $\pm$ 0.8 & 211.4 $\pm$ 1.3 & -- & 266.4 $\pm$ 0.9 & 11.6 $\pm$ 0.5 & 203.2 $\pm$ 0.4 & 19.9 $\pm$ 0.4 \\
	    
	        2 453 662.766 03 & 0.412 71 &  47.2 $\pm$ 0.4 & 173.1 $\pm$ 0.2 & 203.2 $\pm$ 0.4 &  211.4 $\pm$ 0.6 & 162.9 $\pm$ 1.2 & 94.6 $\pm$ 0.3 & 219.9 $\pm$ 1.2 & 31.6 $\pm$ 0.6 & 158.1 $\pm$ 0.5 & 40.5 $\pm$ 0.5 \\

	        $\dots$ & $\dots$ &  $\dots$ & $\dots$ & $\dots$ &  $\dots$ & $\dots$& $\dots$ &  $\dots$ & $\dots$& $\dots$ &  $\dots$ \\
	    
	        $\dots$ & $\dots$ &  $\dots$ & $\dots$ & $\dots$ &  $\dots$ & $\dots$& $\dots$ &  $\dots$ & $\dots$& $\dots$ &  $\dots$ \\
	    
	        $\dots$ & $\dots$ & $\dots$ & $\dots$& $\dots$ &  $\dots$ & $\dots$& $\dots$ &  $\dots$ & $\dots$& $\dots$ &  $\dots$ \\
	    
	        \hline
	    \end{tabular}
	    \end{center}
	    \end{table*}
        \textbf{Note.} The subscription $(1,\textup{abs})$ indicates the primary component and absorption line. The subscription $(1,\textup{emi})$ indicates the primary component and emission line. The subscription $(2,\textup{emi})$ indicates the secondary component and emission line. The orbital phases used in this table are calculated from the ephemeris of \citet{Maxted2007}.
\end{landscape}
\twocolumn

\section{Posterior probability distributions of MCMC fitting parameters.}
\begin{figure*}
	\includegraphics[width=1.1\columnwidth]{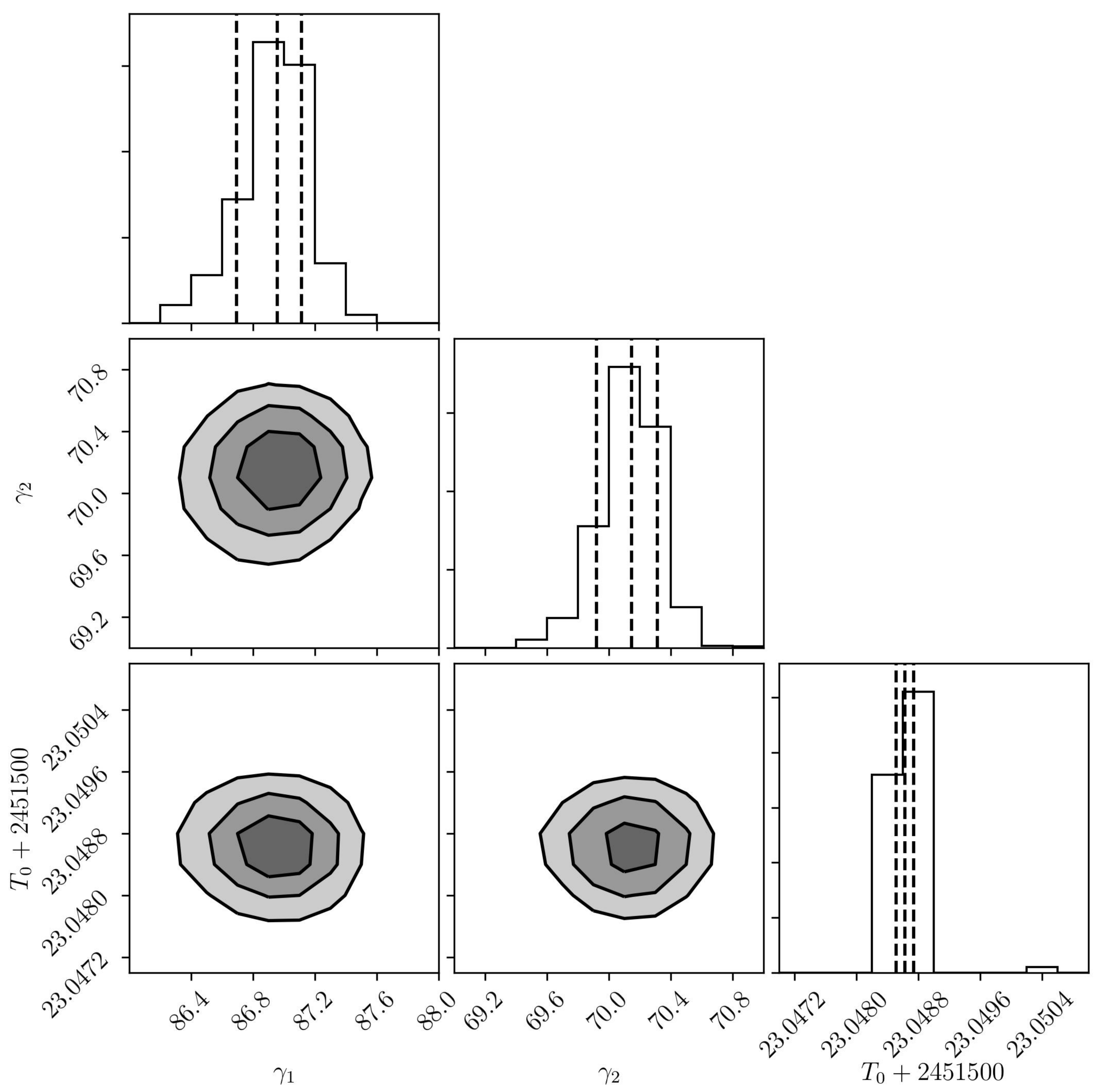}
    \caption{Posterior probability distributions of the radial velocity parameters of RR Cae using the \texttt{PHOEBE} code with the MCMC fitting. The dashed-lines mark the 16th, 50th, and 84th percentiles.}
    \label{fig:param_rv}
\end{figure*}

\begin{figure*}
	\includegraphics[width=1.2\columnwidth]{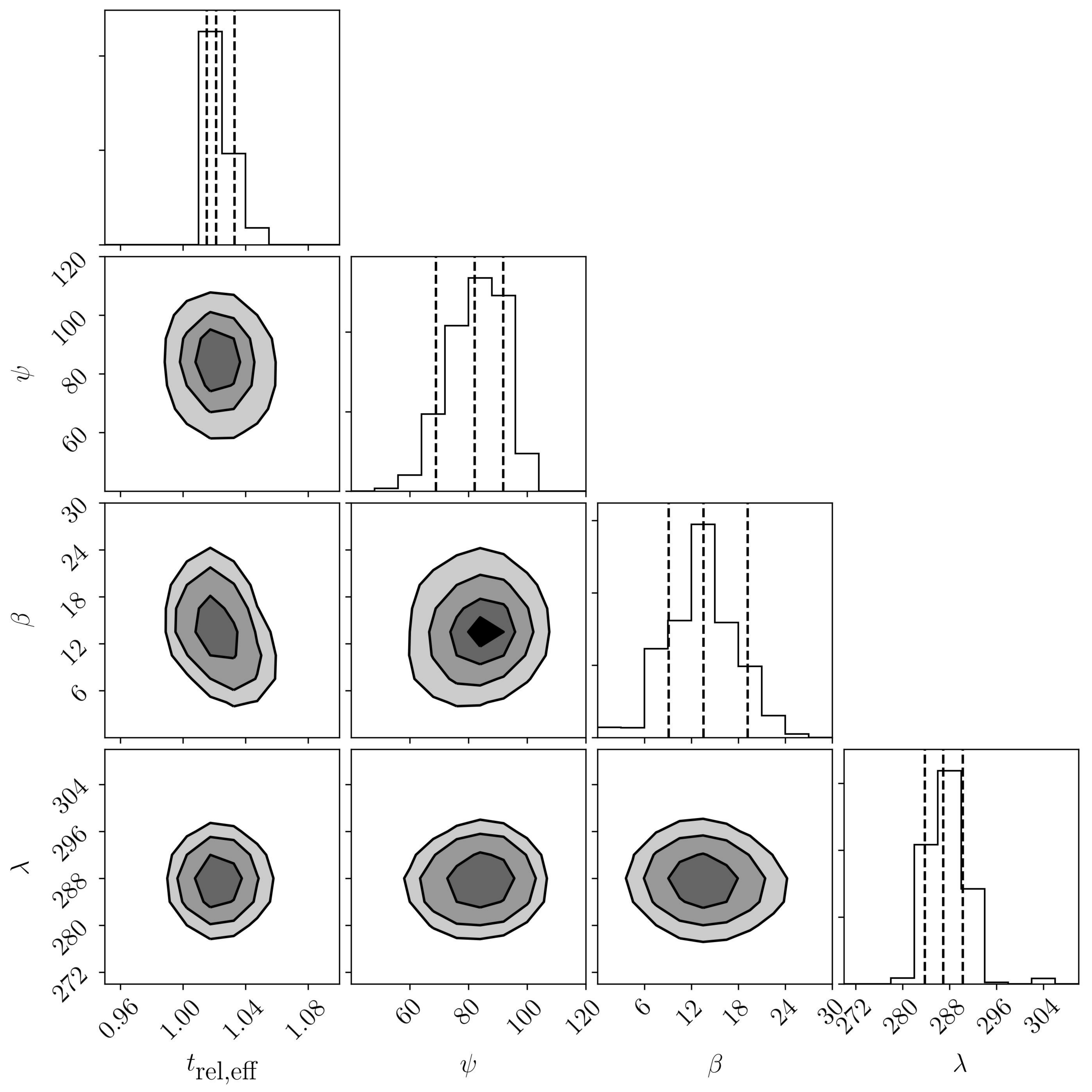}
    \caption{Posterior probability distributions of the spot parameters on the RR Cae TESS light curves in 2018 using the \texttt{PHOEBE} code with the MCMC fitting. The dashed-lines mark the 16th, 50th, and 84th percentiles.}
    \label{fig:param_spot1}
\end{figure*}

\begin{figure*}
	\includegraphics[width=1.2\columnwidth]{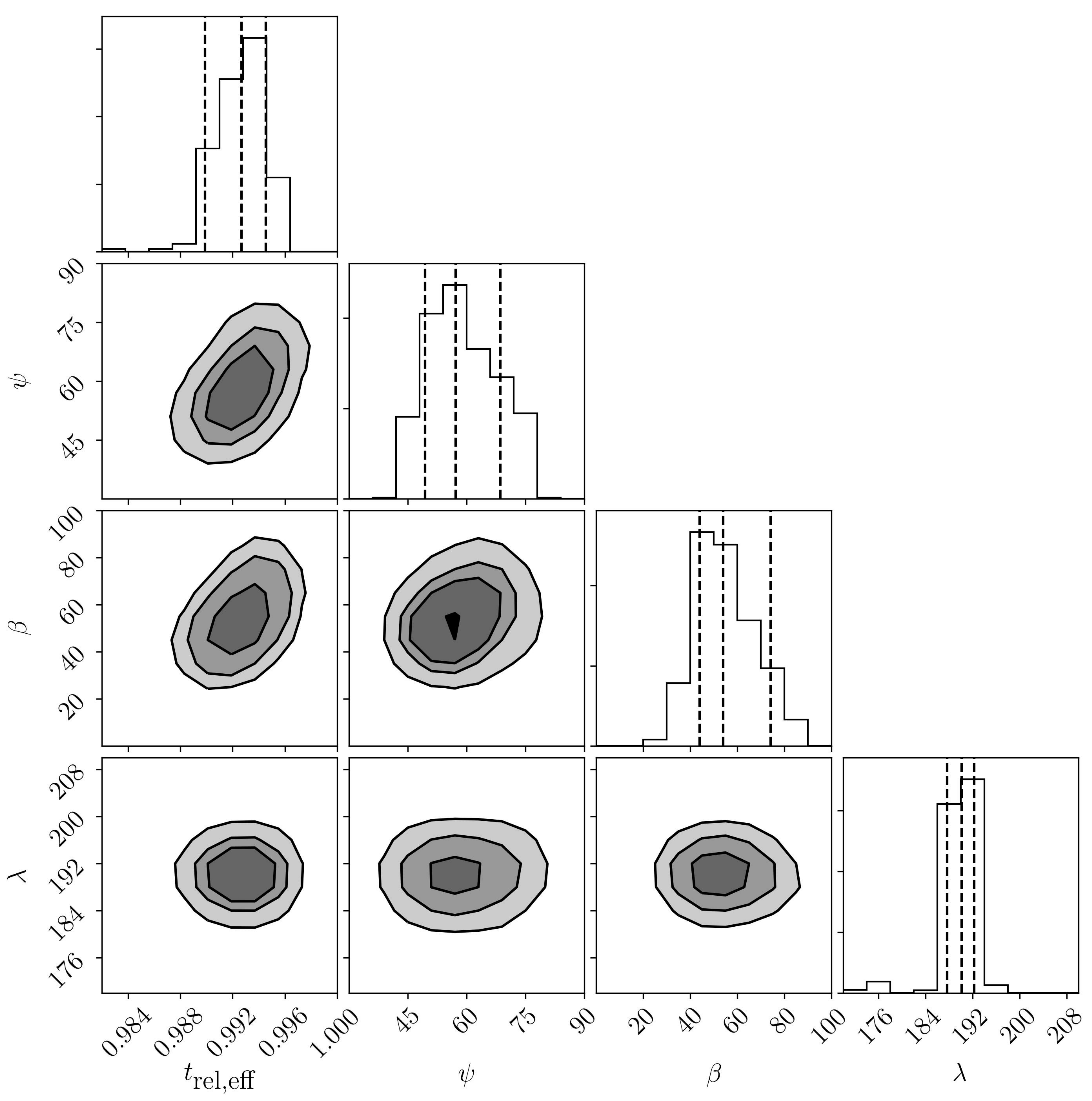}
    \caption{Posterior probability distributions of the spot parameters on the RR Cae TESS light curves in 2020 using the \texttt{PHOEBE} code with the MCMC fitting. The dashed-lines mark the 16th, 50th, and 84th percentiles.}
    \label{fig:param_spot2}
\end{figure*}

\begin{figure*}
	\includegraphics[width=2\columnwidth]{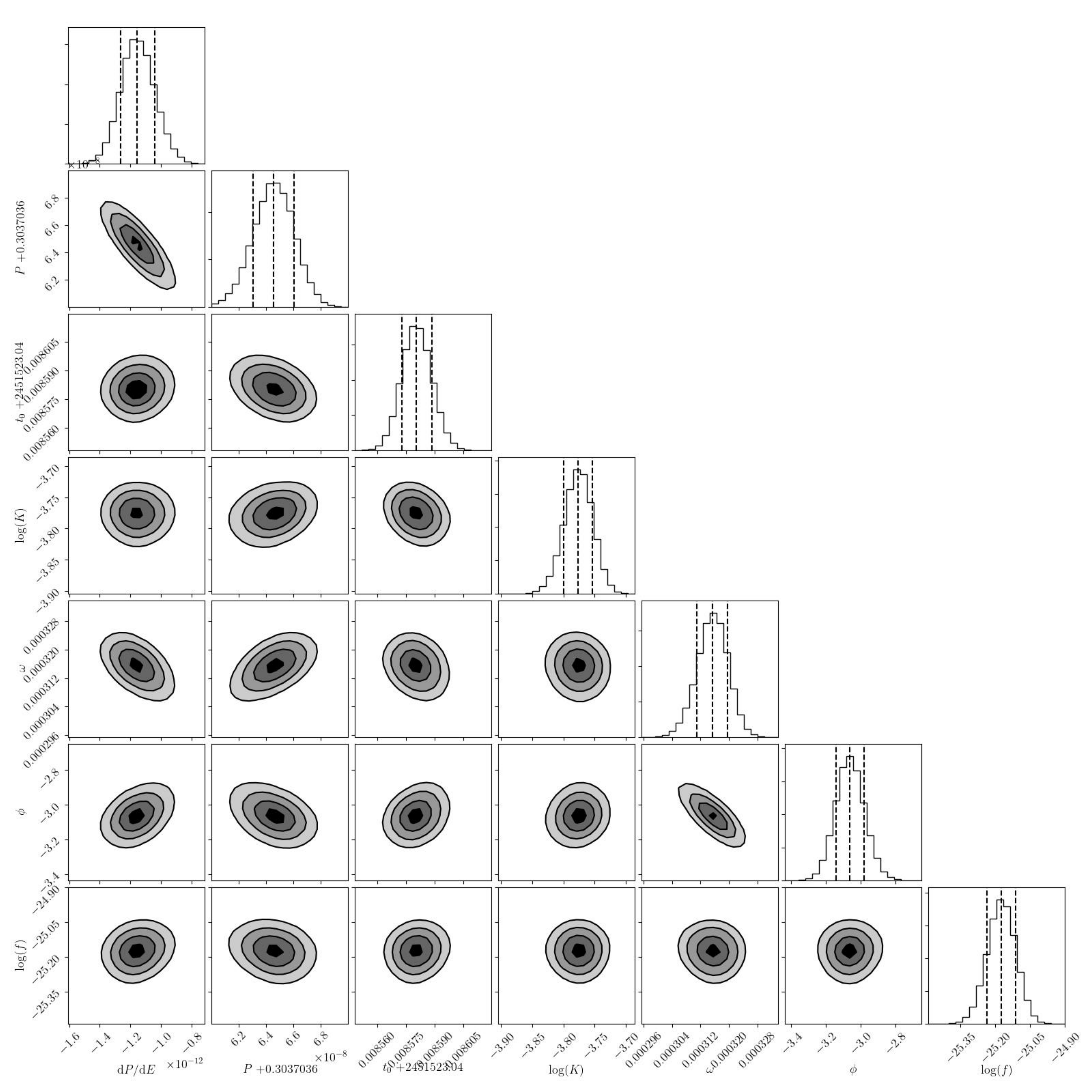}
    \caption{Posterior probability distributions of the O-C MCMC fitting parameters of the one circumbinary object model. The dashed-lines mark the 16th, 50th, and 84th percentiles.}
    \label{fig:ocmcmc}
\end{figure*}

\begin{figure*}
	\includegraphics[width=2\columnwidth]{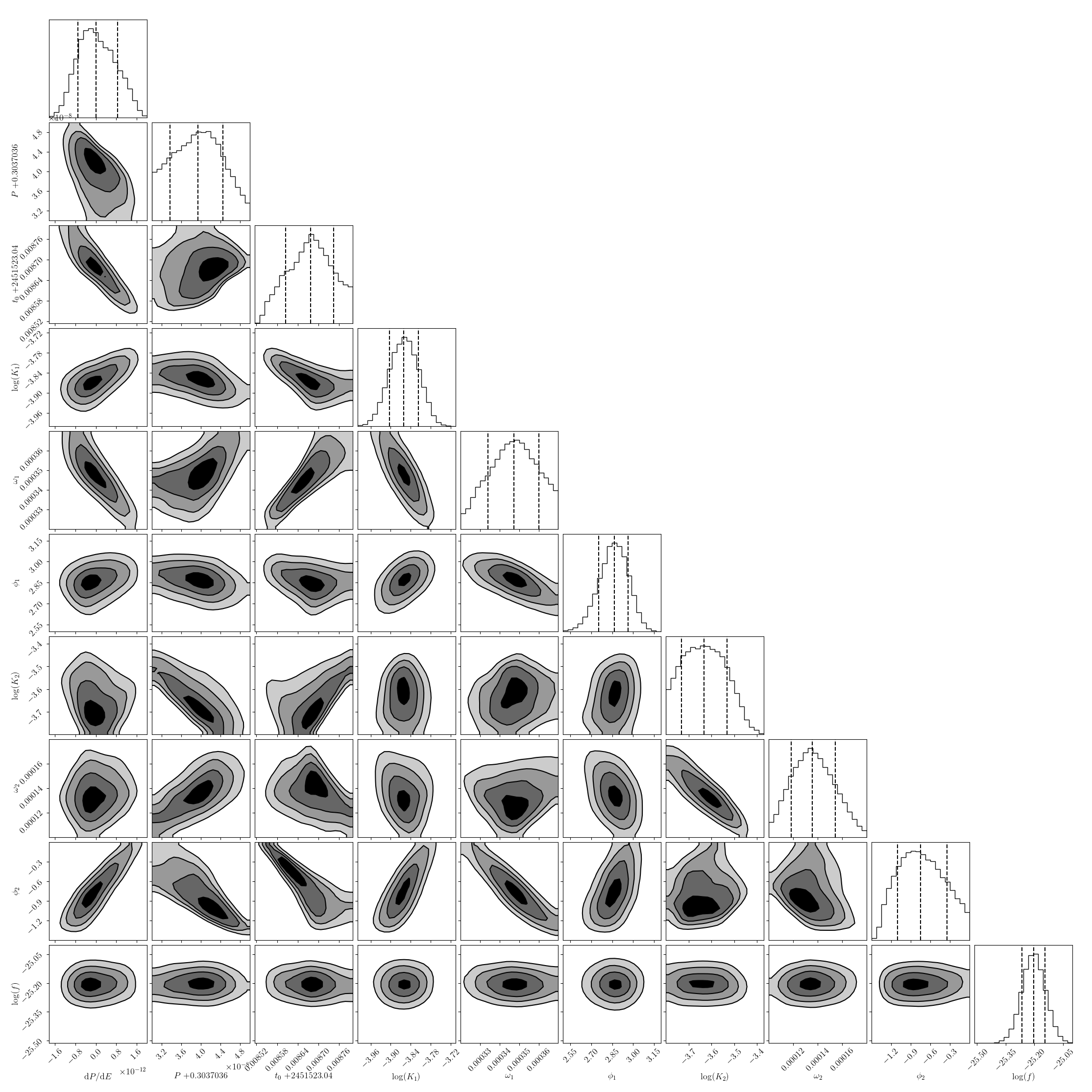}
    \caption{Posterior probability distributions of the O-C MCMC fitting parameters of the two circumbinary objects model. The dashed-lines mark the 16th, 50th, and 84th percentiles.}
    \label{fig:ocmcmc2}
\end{figure*}

\end{document}